\documentstyle[epsf]{mn}  
\newcommand{\bm}[1]{{\bf#1}}

\title{Source-lens clustering effects on the skewness of the lensing
convergence}
\author[T. Hamana et al.]
{Takashi Hamana${}^{1,2}$, St\'ephane T. Colombi${}^{1,3}$, 
Aur\'elien Thion${}^1$, \newauthor Julien E. G. T. Devriendt${}^4$,
Yannick Mellier${}^{1,5}$ and Francis Bernardeau${}^6$\\
$^1$Institut d'Astrophysique de Paris, CNRS, 98bis Boulevard Arago, F75014
PARIS, France\\
$^2$Present address: National Astronomical Observatory of Japan, 
Mitaka, Tokyo 181-8588, Japan\\
${}^3$NIC (Numerical Investigations in Cosmology), CNRS\\
$^4$Nuclear \& Astrophysics Laboratory, University of Oxford, Keble road, 
OX1 3RH Oxford UK\\
${}^5$Observatoire de Paris, DEMIRM, 61 avenue de l'Observatoire,
75014 PARIS, France\\
${}^6$Service de Physique Th\'eorique, C.E.~de Saclay, 91191 Gif sur 
Yvette Cedex, France}
\date{Accepted ....; Received ....; in original form ....}
\pagerange{\pageref{firstpage}--\pageref{lastpage}}
\pubyear{....}
\begin{document}
\maketitle
\label{firstpage}

\begin{abstract}
The correlation between source galaxies and lensing potentials causes
a systematic effect on measurements of cosmic shear statistics, known as
the source-lens clustering (SLC) effect. The SLC effect on the
skewness of lensing convergence, $S_3$, is examined  using a nonlinear
semi-analytic approach and is checked against numerical simulations.
The semi-analytic calculations have been performed in a wide variety
of generic models for the redshift distribution of source galaxies and
power-law models for the bias parameter between the galaxy and dark matter
distributions.  The semi-analytic predictions are tested successfully
against numerical simulations. We find the relative amplitude of the SLC
effect on $S_3$ to be of the order of five to forty per cent. It depends
significantly on the redshift distribution of sources and on the way
the bias parameter evolves.  We discuss possible measurement
strategies to minimize the SLC effects.
\end{abstract}


\section{Introduction}
Recent detections of the cosmic shear signal have opened a new window
to probe the distribution of matter in the Universe, its evolution,
and to test cosmological models (Van Waerbeke et al.~2000; Wittman et
al.~2000; Bacon, Refregier \& Ellis 2000; Kaiser, Wilson \& Luppino
2000; Maoli et al 2001). These detections have been obtained from
relatively small fields so far, which limits the statistical analysis
of the surveys to second order moments, the variance or two-point
correlation function of cosmic shear.  The amplitude of second order
statistics reflects that of density fluctuations and roughly scales as
$\sigma_\gamma \propto\Omega_{\rm m}^{0.6-0.8}\sigma_8$ at large scale
(Bernardeau, Van Waerbeke \& Mellier 1997, hereafter BvWM97) and
$\sigma_\gamma \propto\Omega_{\rm m}^{0.6-0.8}\sigma_8^{1.2-1.3}$ at
small scale (Jain \& Seljak 1997; Maoli et al. 2001).  On the other
hand, the skewness (a third order statistic) 
of lensing convergence is known to be sensitive to
$\Omega_{\rm m}$,  almost independently of $\sigma_8$ (BvWM97).
Therefore, combined analysis of the skewness and the variance will
provide precious constraints on both values of $\Omega_{\rm m}$ and
$\sigma_8$.  As a consequence,  skewness detection and measurement  is
one of main goals of on-going wide field cosmic shear surveys such as
the DESCART project\footnote{For more information about DESCART
project, see http://terapix.iap.fr/Descart/}.

Cosmic shear statistics have been studied analytically (see Mellier
1999 and Bartelmann \& Schneider 2001 for reviews and references
therein) as well as numerically (Jain, Seljak \& White 2000; White \&
Hu 2000).  The skewness of lensing convergence was first calculated by
BvWM97 based on a quasi-linear perturbation theory approach.  It has
been, however, recognized that this approach is not robust enough to
provide accurate predictions for the value of the skewness over the whole
available dynamic range. In particular,  the  two following points
have to be addressed and carefully included in the calculations: (i)
Nonlinear growth of the density field: numerical studies show that
nonlinear growth enhances skewness especially at angular scales
smaller than one degree (Jain et al.~2000; White \& Hu 2000; Van
Waerbeke et al.~2001b).  (ii) Source clustering: Bernardeau (1998)
(hereafter B98) pointed out that correlations between source galaxies
and  lensing potential reduce skewness amplitude.  B98 underlined that
this effect is sensitive to the redshift distribution  of sources.

The purpose of this paper is to examine the effect of source
clustering (SLC) on measurements of the skewness of lensing
convergence.  Special attention is payed to its dependence on  the
redshift distribution of sources and on evolution of the bias relation
between matter and galaxy distribution. Since the redshift distribution
of faint galaxies is uncertain and little is known about the bias,
this paper does not aim at making accurate predictions for the
amplitude of SLC effect in real cosmic shear surveys.  Our objective
is to estimate its  magnitude in order to propose  strategies that
minimize its effects.

We basically follow the perturbation theory approach first developed
by B98 but generalize it in two ways: (i) we take into account the
effects of nonlinear evolution of the density field, adopting the
nonlinear semi-analytic ansatz developed  by Jain \& Seljak (1997) and
Van Waerbeke et al.~(2001b); (ii) we allow a possible redshift
dependence of the bias parameter, $b(z)=b_0(1+z)^\gamma$ and examine
the cases $\gamma=0$ to 2.  Moreover we consider three cosmological
Cold Dark Matter family models (CDM),  two flat models with and
without cosmological constant and an open model, and 12 different
models for the source distribution which cover a wide range of mean
redshift and width for the distribution.

Finally, for the first time the accuracy of semi-analytic predictions
for the SLC effects on the skewness is tested against numerical
simulations in standard CDM model. 

The outline of this paper is as follows.  In section 2, the physical
mechanism of SLC is described.  In section 3, an expression for the
skewness of lensing convergence is presented that takes both SLC and
nonlinear evolution of the density field into account.  In section 4,
our models are described.  Results of the semi-analytic approach are
presented in section 5.  In section 6, semi-analytic predictions are
tested against numerical simulations.  We summarize and discuss our
conclusions in section 7.  The derivation of the convergence skewness
in the presence of the SLC is presented in Appendix A.  The $N$-body
data sets and the ray-tracing method used for this work are described
in Appendix B.  In Appendix C, details of the procedure to generate
mock galaxy catalogues are presented.

\section{What is the SLC effect?}

\begin{figure}
\begin{center}
\begin{minipage}{8.5cm}
\epsfxsize=8.5cm \epsffile{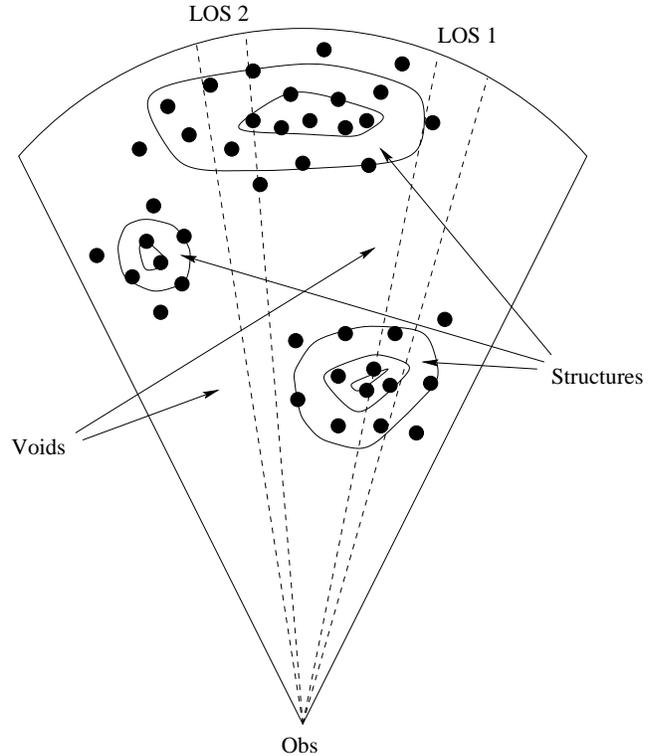}
\end{minipage}
\end{center}
\caption{An illustration of the correlation between the gravitational
potential (contour lines) and the population of sources (denoted by
filled circles).  }
\label{fig:illust}
\end{figure}

The SLC effect discussed in this paper comes to light because of the
conjunction of three circumstances, namely: (i) source galaxies are
not randomly distributed in the sky but are correlated; (ii) the
source galaxy distribution traces somehow  the matter field; (iii) the
redshift distribution of source galaxies is rather broad. The width
of the distribution
depends on source selection criterion, and generally, the distribution
of source galaxies overlaps with the distribution of lensing
structures, so source galaxies are somehow correlated with
the lensing potential.  This correlation causes systematic effects on
measurements of cosmic shear, that may be illustrated as follows.
Figure \ref{fig:illust} shows a distribution of
sources (denoted by filled circles) and the gravitational potential
(contour lines). For line-of-sight 1 (LOS 1), the distant galaxies
are lensed by the gravitational potential located at an intermediate
distance and thus have a high {\it positive} lensing
convergence\footnote{The lensing convergence is not a direct
observable but is obtained via a convergence reconstruction technique
(Van Warbeke, Bernardeau \& Mellier 1999) or the aperture mass
(Schneider et al.~1998) from a lensing shear map.}.  This high signal
is reduced by the excess of foreground sources bound to the foreground
gravitational potential which, in contrast, has a low lensing.  On the
other hand, for a line-of-sight 2 (LOS 2), distant sources are lensed
by the foreground void and thus have a {\it negative} lensing
convergence.  This negative signal is amplified because of the lack of
foreground sources in the void.  Accordingly, the probability
distribution function of the lensing convergence, which is skewed
toward a high value in absence of the SLC (e.g., BvWM97, Jain et
al.~2000) becomes more symmetric than for the case of a random
distribution of source galaxies.  As a result, the amplitude of
skewness of lensing convergence drops.

As was pointed out by B98, there is another possible effect caused by
intrinsic clustering of source galaxies. The average distance of
sources and their sky density may indeed vary from one direction to
another, which can
cause additional systematic effects on the cosmic shear statistics 
(these effects was included in the definition of source clustering used
by B98).
It was pointed out by B98 and Thion et al.~(2001) that this effect on
the convergence skewness is in general very small.  Therefore, in this
analysis, we do not take it into account for the analytical
calculations presented in next section, although it will be obviously
present in the numerical experiments discussed in \S~6.

\section{The perturbation theory approach}
\subsection{The quasi-linear regime}

The expressions for the skewness of lensing convergence and the
correlation term due to SLC were first derived by BvWM97 and B98 in
the framework of perturbation theory.   In this subsection, we only
summarize expressions which are directly relevant to this
paper. Detailed derivations  and notations are given in Appendix~A.

In the presence of the SLC, the skewness parameter, defined by
$S_3(\theta)=\langle \kappa_\theta^3\rangle / V_\kappa^2(\theta)$,
consists of two terms; namely, one arises
from the quasi-linear theory and the other from SLC:
\begin{eqnarray}
\label{S3}
S_3(\theta) = S_3^{\rm q.l.}(\theta)+ S_3^{\rm slc}(\theta),
\end{eqnarray}
with
\begin{eqnarray}
\label{S3ql}
S_3^{\rm q.l.}(\theta) &=& {{\langle \kappa_\theta^3 \rangle^{q.l.}}
\over {V_\kappa^2(\theta)}}\nonumber\\  &=& {6 \over
{V_\kappa^2(\theta)}}  \left({{H_0}\over c}\right)^3 \int_0^{\chi_H} d
\chi\, w^3(\chi) \nonumber\\ && \times \left[ {6\over7}
I_0^2(\chi,\theta)+{1\over4} I_0(\chi,\theta) I_1(\chi,\theta)\right],
\end{eqnarray}
\begin{eqnarray}
\label{S3sc}
S_3^{\rm slc}(\theta) &=&  {{\langle \kappa_\theta^3\rangle^{\rm slc}}
\over {V_\kappa^2(\theta)}}\nonumber\\ &=& {{9 \Omega_{\rm m}} \over
{V_\kappa^2(\theta)}} \left({{H_0}\over c}\right)^3 \int_0^{\chi_H} d
\chi\, n_s(\chi)  b(\chi) w(\chi) I_0(\chi) \nonumber\\ && \times 
\int_0^{\chi} d \chi'\, g(\chi',\chi) w(\chi')  I_0(\chi')\nonumber\\
&&-{6 \over {V_\kappa(\theta)}} {{H_0} \over c} \int_0^{\chi_H} d
\chi\,  n_s(\chi) b(\chi) w(\chi) I_0(\chi).\nonumber\\
\end{eqnarray}
In these equations,
\begin{itemize}
\item $\theta$ is the size of the smoothing window,
\item $H_0/c$ is the Hubble constant in speed of light units,  
\item $V_k(\theta)$
is the variance of the lensing convergence [given by eq.~(\ref{variance})],
\item $\chi$ denotes the radial comoving distance (and $\chi_H$ corresponds
to the horizon), 
\item $w(\chi)$ is the lensing efficiency function
[eq.~(\ref{effeciency})],  
\item functions $I_0(\chi,\theta)$ and
$I_1(\chi,\theta)$ depend on the power-spectrum of matter density
fluctuations through eqs.~(\ref{eq:ap:I0}) and (\ref{I1}),
\item $\Omega_{\rm m}$ is the matter density in units of the critical
density, 
\item $n_s[\chi(z)]$ the average number density of sources,
\item function $g(\chi_l,\chi_s)$ is given by eq.~(\ref{eq:ap:g}), 
\item $b(\chi)$
is the linear biasing function between the galaxy and the matter
density contrast. 
\end{itemize}
\subsection{Nonlinear regime}

For the variance of the lensing convergence,  the effect of nonlinear
evolution of the density power spectrum can be included by replacing
the linear power spectrum (which enters the above expressions through
$V_\kappa(\theta)$, $I_0(\chi)$ and $I_1(\chi)$, see Appendix A for
their explicit expressions)  with the nonlinear  power spectrum, i.e.,
$P_{\rm lin}(a,k) \rightarrow P_{\rm NL}(a,k)$ (Jain \& Seljak 1997).
We use the fitting formula of nonlinear power spectrum  given by
Peacock and Dodds  (1996).  This semi-analytic approach has been
tested against ray-tracing  simulations, and a good agreement between
the numerical results and the  semi-analytic predictions was found
(Jain et al 2000; White and Hu 2000).

In the framework of perturbation theory, all density contrasts needed
for  the calculation of the skewness correction term [equation
(\ref{S3sc})] correspond to linear order (see B98 for details).  This
comes from the fact that the quantity to be computed, an angular average
of projected fourth moments, is given by the product of two two-point
correlations: its intrinsic connected part has a negligible
contribution because of the projection effects. The incorporation of
the nonlinear effects is then straightforward. As for the variance, it
amounts to formally replacing the linear power spectrum with the
nonlinear one.

The semi-analytic calculation of the skewness in the nonlinear regime
was developed by Van Waerbeke et al.~(2001b). It is based on the
fitting formula of the density bispectrum by  Scoccimarro \& Couchman
(2000) and is given by
\begin{eqnarray}
\label{S3nl}
S_3^{\rm n.l.}(\theta)  &=& {1\over{V_\kappa^2(\theta)}}{6 \over {(2
\pi)^4}} \left({{H_0}\over c}\right)^3  \int_0^{\chi_{\rm max}} d
\chi\, w^3(\chi) \nonumber\\ &&  \times \int d^2 \bm{k_1}\,
P_{\rm NL}(k_1)  W[k_1 f(\chi) \theta] \nonumber\\ && \times \int d^2
\bm{k_2}\, P_{\rm NL}(k_2)  W[k_2 f(\chi) \theta]\nonumber\\ &&
\times W[|\bm{k_1}+\bm{k_2}| f(\chi) \theta]  F_2^{\rm
eff}(\bm{k_1},\bm{k_2}),
\end{eqnarray}
with
\begin{eqnarray}
\label{F2eff}
F_2^{\rm eff}(\bm{k_1},\bm{k_2})  &=& {5 \over 7}
a(n,k_1)a(n,k_2)\nonumber\\ && +{1 \over 2} b(n,k_1)
b(n,k_2){{\bm{k_1}\cdot\bm{k_2}} \over {k_1 k_2}}  \left({{k_1} \over
{k_2}}+{{k_2} \over {k_1}} \right)\nonumber\\ && +{2 \over 7}
c(n,k_1)c(n,k_2) {({\bm{k_1}\cdot \bm {k_2}})^2 \over  {|k_1|^2
|k_2|^2}}.
\end{eqnarray}
Here, the notations are as follows:
\begin{itemize}
\item $k=\vert \bm{k} \vert$, 
\item $f(\chi)$ denotes the comoving
angular diameter distance (see Appendix A), 
\item $W$ is the Fourier
transform of the smoothing window, 
\item functions $a(n,k)$, $b(n,k)$ and $c(n,k)$
depend on the  effective power spectral index $n$ at scale $k$
(explicit expressions are given in  Scoccimarro \& Couchman 2000; see
also Van Waerbeke et al.~2001b).  
\end{itemize}
It should be noted that,  since the
bispectrum fitting formula is constructed via the dark matter
bispectrum measured from only one $N$-body simulation data set, there
is about a 10-20 per cent uncertainty in the fitting formula. This is
mainly a cosmic variance effect (Van Waerbeke et al.~2001b).

\section{Models}
\subsection{Cold dark matter models (CDM)}

\begin{table}
\caption{Cosmological parameters for the models considered in this
paper. $\Omega_{\rm m}$ is the matter density in units of the critical
density and similarly for the cosmological constant
$\Omega_{\lambda}$; $h\equiv H_0/100$ is the Hubble constant in units
of $100$ km s$^{-1}$ Mpc$^{-1}$; $\sigma_8$ is the variance of the dark matter
distribution in a sphere of radius $8\,h^{-1}$ Mpc.}
\label{table:cosmo}
\begin{center}
\begin{tabular}{ccccc}
\hline Model & $\Omega_{\rm m}$ & $\Omega_\lambda$ & $h$ & $\sigma_8$
\\ \hline SCDM & 1.0 & 0.0 & 0.5 & 0.6 \\ OCDM & 0.3 & 0.0 & 0.7 &
0.85 \\ $\Lambda$CDM & 0.3 & 0.7 & 0.7 & 0.9 \\ \hline
\end{tabular}
\end{center}
\end{table}

We discuss three Cold Dark Matter (CDM) models, a flat model with
($\Lambda$CDM) and without cosmological constant (SCDM) and an open
model (OCDM), using galaxy cluster abundances to normalize the
power-spectrum (Eke, Cole \& Frenk, 1996; Kitayama \& Suto 1997) and
the formula of Bond \& Efstathiou (1984) for the transfer function.
The parameters of the models are listed in Table \ref{table:cosmo}.

\subsection{Redshift distribution of source galaxies}

We assume that $n_s(z)$ takes the form,
\begin{eqnarray}
\label{ns}
n_s(z) dz =  { {dz \beta} \over {z_\ast
\Gamma[(1+\alpha)/\beta]}}  \left( {z\over {z_\ast}} \right)^\alpha
\exp \left[-\left({z\over{z_\ast}} \right)^\beta \right],
\end{eqnarray}
where $\Gamma(x)$ is the Gamma function.

\begin{table}
\caption{Parameters in $n_s(z)$.}
\label{table:ns}
\begin{center}
\begin{tabular}{cccccc}
\hline Model & $\langle z \rangle$ & $\Delta z$ & $\alpha$ & $\beta$ &
$z_\ast$ \\ \hline A1 & 1.2 & 0.572 & 2 & 1.5 & 0.798\\ A2 & 1.2 &
0.456 & 3 & 1.8 & 0.813\\ A3 & 1.2 & 0.297 & 5 & 3.0 & 1.01\\ A4 & 1.2
& 0.182 & 8 & 6.0 & 1.18\\ B1 & 1.5 & 0.866 & 2 & 1.0 & 0.500\\ B2 &
1.5 & 0.618 & 3 & 1.5 & 0.812\\ B3 & 1.5 & 0.400 & 5 & 2.5 & 1.11\\ B4
& 1.5 & 0.244 & 7 & 6.0 & 1.51\\ C1 & 0.9 & 0.429 & 2 & 1.5 & 0.598\\
C2 & 0.9 & 0.342 & 3 & 1.8 & 0.610\\ C3 & 0.9 & 0.240 & 5 & 2.5 &
0.667\\ C4 & 0.9 & 0.136 & 8 & 6.0 & 0.884\\ \hline
\end{tabular}
\end{center}
\end{table}

We explore 12 models for the shape of the distribution.
The parameters in each model are listed in Table \ref{table:ns}.
The average redshift is $\langle z \rangle$ = 1.2, 1.5 and 0.9 for
models A1-4, B1-4 and C1-4, respectively.  We characterize the width
of the distribution by the root-mean-square, $\Delta z$, which varies
within a factor of $\simeq$3.2, 3.5 and 3.2 in models A1-4, B1-4 and
C1-4, respectively.  Note that only model A1 matches  roughly the
observed redshift distribution of galaxies in current cosmic shear
detections (Van Waerbeke et al.~2000).  However, to keep our approach
as general  as possible, we still use a reasonably large parameter
range for the possible shapes of the distributions.

\begin{figure}
\begin{minipage}{8cm}
\begin{center} 
\epsfxsize=8cm \epsffile{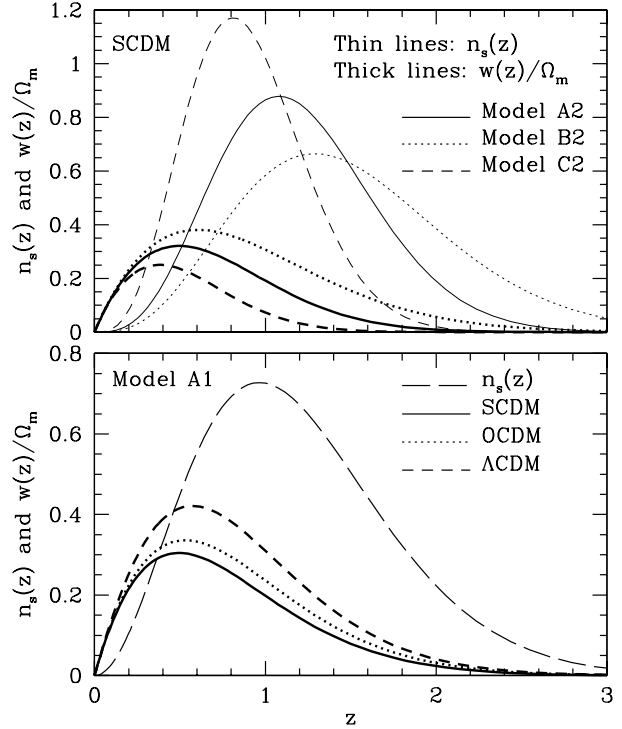}
\end{center}
\end{minipage}
\caption{The redshift distributions of sources, $n_s(z)$, and the
corresponding lensing efficiency functions divided by the density
parameter, $w(z)/\Omega_{\rm m}$, as functions of redshift.  {\it Top
panel:} for three source distribution models in SCDM model.  {\it
Bottom panel:} for A1 model in three cosmologies.}
\label{fig:ns_w}
\end{figure}

Figure \ref{fig:ns_w} shows the redshift distribution of sources and
the corresponding lensing efficiency as functions of
redshift. In the top panel, SCDM is supposed, and various models for
galaxy number counts are taken.  In the bottom panel, model A1 is assumed
for number counts, and various cosmologies are considered.  Roughly
speaking, the amplitude of SLC is controlled by the amplitude of
overlapping between the population of sources [$n_s(z)$] and that of
lenses [which is very closely related to $w(z)/\Omega_{\rm m}$].  It
is important to keep in mind that the normalized efficiency function
$w(z)/\Omega_{\rm m}$ increases in order of SCDM, OCDM and
$\Lambda$CDM.

\subsection{Model for the bias}
We assume that the bias between the galaxy and the matter distribution
is linear and takes a power-law form as a function of redshift, i.e.,
\begin{eqnarray}
\label{bias}
b(z)=b_0(1+z)^\gamma.
\end{eqnarray}
We examine three cases, $\gamma=0$, 1 and 2, and we shall take $b_0=1$. 
Since, so far, little is
known about a realistic description of the bias, we adopted this model
for its simplicity and the wide possible range of possibilities it
nevertheless covers.  Numerical studies of dark matter clustering
combined with  measurements of two-point correlation function in
galaxy catalogues  suggested that $b_0$ is close  to unity (e.g.,
Jenkins et al.~1998).

\section{Results}

\begin{figure}
\begin{center}
\begin{minipage}{8cm}
\epsfxsize=8cm \epsffile{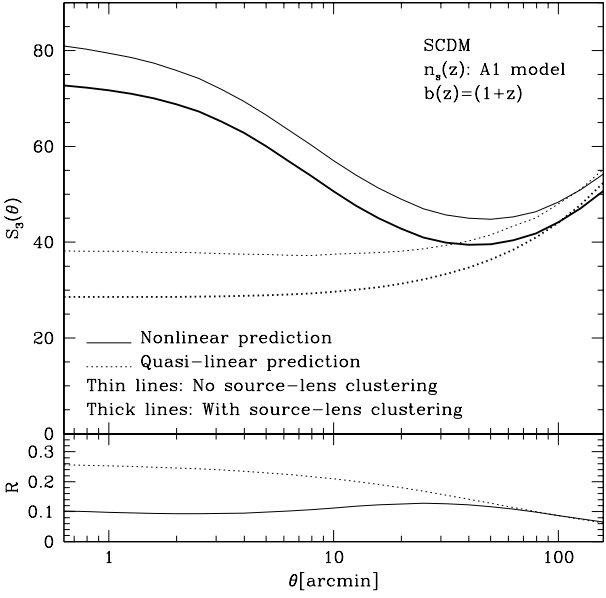}
\end{minipage}
\end{center}
\caption{Upper panel: predicted skewness of the lensing convergence
for A1 and SCDM, with and without source  clustering effect taken into
account.  Lower panel: the ratio $R$ as defined in equation
(\ref{R}).  }
\label{fig:li_nl}
\end{figure}

Let us introduce the parameter which characterizes the amplitude of
the SLC effects defined by
\begin{eqnarray}
\label{R}
R=- {{S_3^{\rm slc}} \over {S_3}},
\end{eqnarray}
where $S_3^{\rm slc}$ is the correction brought by SLC and $S_3$ is the
skewness in the case of the absence of SLC ($S_3^{\rm q.l.}$ and
$S_3^{\rm n.l.}$ for quasi-linear and nonlinear computation, respectively).

Figure \ref{fig:li_nl} shows $S_3$ (upper panel), with and without
taking the SLC effect into account, and $R$ (lower panel) as a
function of $\theta$ for the A1 model and SCDM.
Nonlinear effects on the skewness are discussed in detail in  Van
Waerbeke et al.~(2001b).  It should be noted that nonlinear growth of
the density field enhances the skewness significantly at scales below
1 degree,  so the SLC correction term remains relatively  small because of
cancellations between the numerator  and the denominator in the first
line of equation ({\ref{S3sc}) (see also Appendix A for the explicit
expressions for $V_\kappa$ and $S_3^{\rm slc}$).  As a consequence, $R$
decreases significantly  when
$\theta < $ 20-30 arcmin.  It should be also noted that for $\theta
\ga 100$ arcmin, where nonlinear effects can be safely neglected, SLC
effect is reduced while $\theta$ increases.  
This is due to the change in the slope of the density power spectrum
occurring when the spatial smoothing scale $f(\chi)\theta$ at $z=0.3\sim0.8$, 
where the most of the lensing contribution comes from, 
is of the order of $10(\Omega_{\rm m} h^2)^{-1}$Mpc.

\begin{figure}
\begin{center}
\begin{minipage}{8cm}
\begin{center} 
\epsfxsize=8cm \epsffile{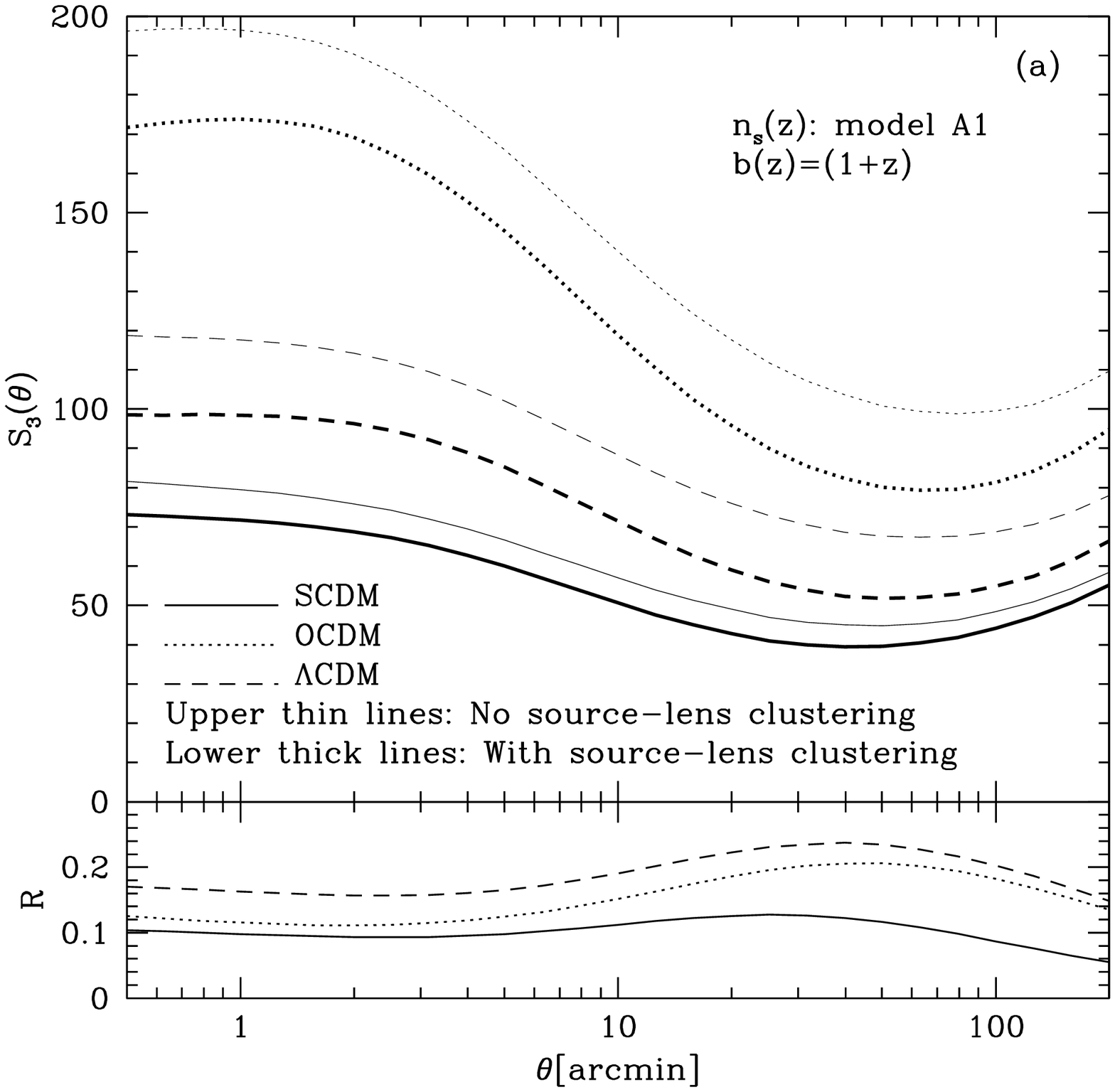}
\end{center}
\end{minipage}
%
\begin{minipage}{8cm}
\begin{center} 
\epsfxsize=8cm \epsffile{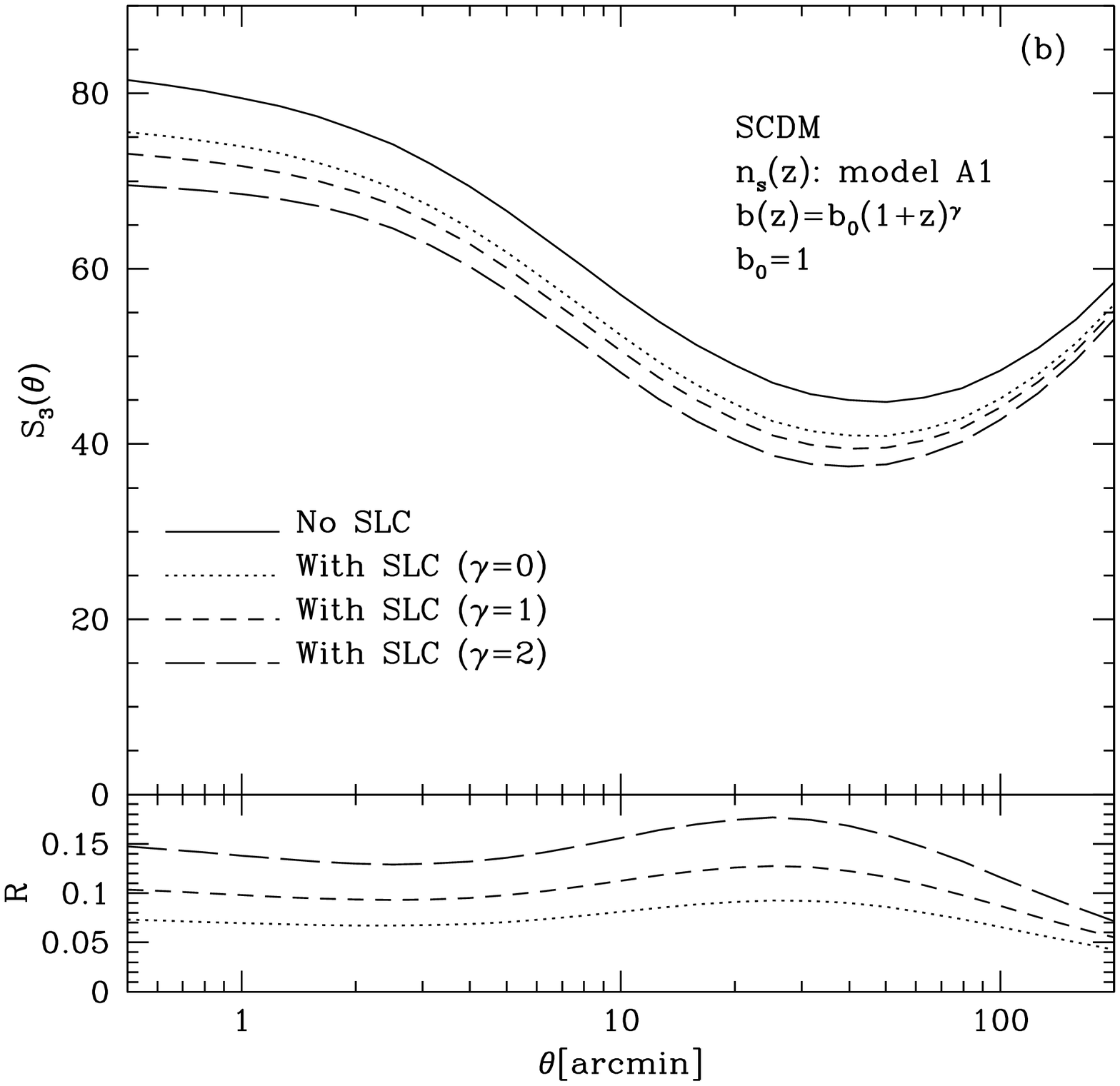}
\end{center}
\end{minipage}
\end{center}
\caption{$S_3$ and $R$ [equation (\ref{R})] with and without the
source  clustering effect taken into account as functions of
scale. The nonlinear ansatz is used.  Cosmology and source
distribution models are denoted in each plot: (a) three different
cosmological models are considered; (b) three different bias evolution
models are considered.}
\label{fig:s3_rat}
\end{figure}

Let us now discuss the theoretical predictions that take into account 
nonlinear effects. 
Figure \ref{fig:s3_rat} shows $S_3$ and $R$ for three cosmologies (top
panel) and three bias evolution models (bottom panel).  The top panel
clearly shows that it is essential to take SLC 
into account  to put constraints on values of  $\Omega_{\rm m}$
determined from $S_3$.  Figure \ref{fig:s3_rat} also suggests
that SLC is more important for low than for high density models.
This is explained by the fact that the efficiency function is larger
in the first than in the second case, as illustrated by Figure
\ref{fig:ns_w}.  The bottom panel of Figure \ref{fig:s3_rat} also shows 
that the SLC effect increases  with strength of evolution in bias with
redshift.  This is a natural consequence of the fact that the SLC effect is
caused by the correlation between the lensing potential (the matter
distribution) and the distribution of source galaxies.  Finally, note
that for $\theta \la 10$ arcmins, the relative SLC effect is nearly
independent of scale.

Figures \ref{fig:li_nl} and \ref{fig:s3_rat} show that the relative SLC
effect, $R$, peaks around $\theta=$30-60 arcmin.  One might
wonder what should be the ideal smoothing scale for measuring the
skewness while reducing as much as possible the SLC effect:  should it
be larger or smaller than the peak position? To answer this
question properly, one also has to consider signal-to-noise ratio, $S/N$.
Typically, signal-to-noise in $S_3$ is expected to decrease with
$\theta$ due to the finite area covered by the survey.  Van
Waerbeke, Bernardeau \& Mellier (1999) numerically investigated  the
efficiency of weak lensing surveys, taking into account both this effect
and the noise due to intrinsic ellipticity of the source
galaxies.  Figures 8 and 10 of their paper indicate that it might be
difficult to  detect the skewness with $S/N>1$ at smoothing scales
larger then  60 arcmin, even with a wide field survey covering
$10^{\rm o}\times10^{\rm o}$.  This suggests that the best choice for the
smoothing scale, keeping both SLC effects  low and a good
signal-to-noise ratio, should be $\theta$ of order of 1
arcmin.\footnote{It  should be however noted that to break the
degeneracy between cosmological parameters, one still has to measure
cosmic shear statistics at linear scales, i.e.~$\theta>1$ degree (Jain
\& Seljak 1997).}

\begin{figure}
\begin{center}
\begin{minipage}{8.2cm}
\epsfxsize=8.2cm \epsffile{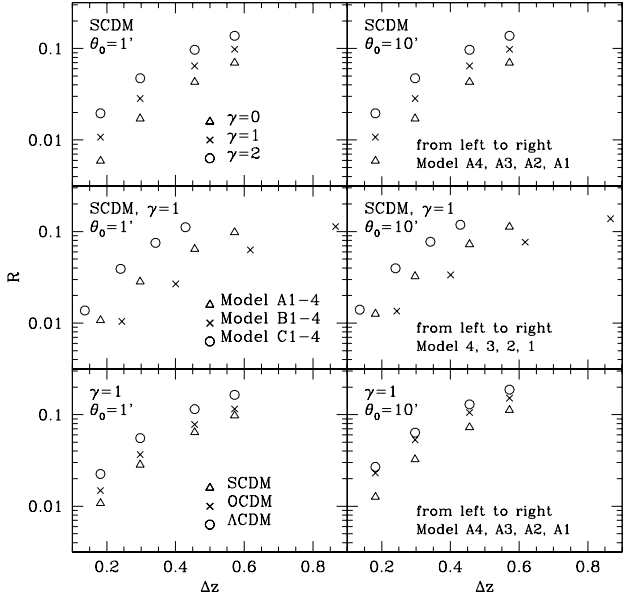}
\end{minipage}
\end{center}
\caption{$R$ as a function of $\Delta z$ as predicted from our
semi-analytic model. The smoothing scales are  1 arcmin and 10 arcmin
for left and right panels, respectively.  Top panels are for three
bias models in SCDM, A1-4, middle panels are for all 12 source
distribution models in SCDM $\gamma=1$ case, and bottom panels are for
three cosmological models in $\gamma=1$ A1-4 cases.  }
\label{fig:rat_dz}
\end{figure}

Figure \ref{fig:rat_dz} shows $R$ for $\theta=1$ arcmin (left panels)
and 10 arcmin (right panels) as a function of the source redshift
distribution width, $\Delta z$.  A comparison of the left and right panels
confirms a visual inspection of Fig.~\ref{fig:s3_rat},  namely that $R$
is fairly  insensitive to $\theta$ in the scaling  regime considered,
$\theta \la 10$ arcmin.  The top and bottom panels indicate that effects
of cosmology and bias evolution model on the amplitude of parameter
$R$ are significant, but the shape  of $R$ as a function of $\Delta z$
remains fairly stable.  Note furthermore that in the middle panels,
models with the same mean source redshift form sequences  in
the $R$-$\Delta z$ plane with very similar slopes at fixed $\Delta z$.

\begin{figure}
\begin{center}
\begin{minipage}{8cm}
\epsfxsize=8cm \epsffile{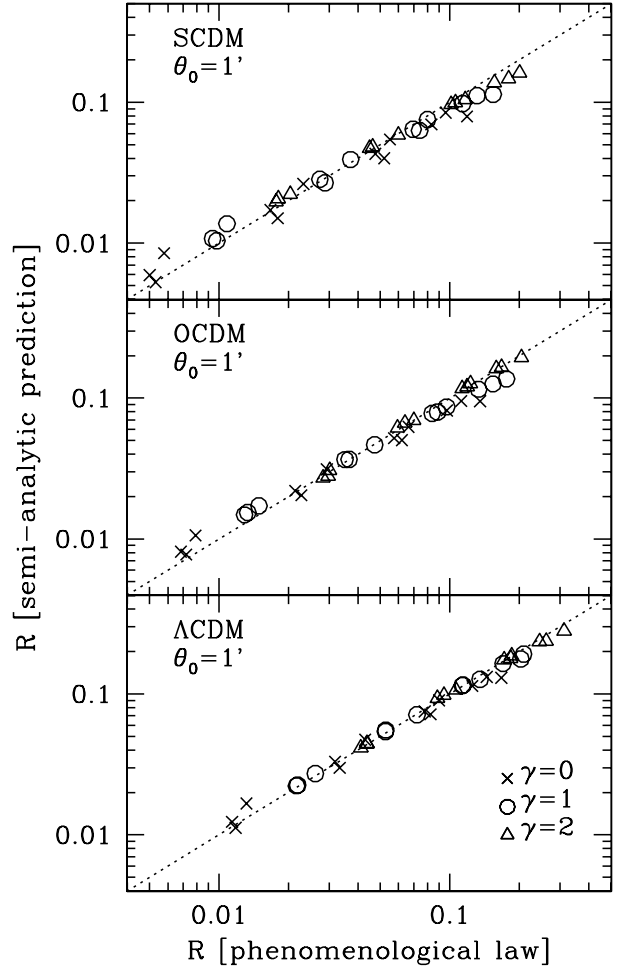}
\end{minipage}
\end{center}
\caption{$R$ computed by the semi-analytic formula versus that derived from 
the phenomenological law [eq.~(\ref{fp})]. 
The parameters $(A,B,C)$ for each case are summarized in 
Table \ref{table:fp}.}
\label{fig:rat_fp}
\end{figure}

\begin{table}
\caption{Parameters of the phenomenological law [eq. (\ref{fp})] derived from
semi-analytic calculations}
\label{table:fp}
\begin{center}
\begin{tabular}{cccccccc}
\hline
 &  & \multicolumn{3}{c}{$\theta=1'$} &
\multicolumn{3}{c}{$\theta=10'$}\\  
 & $\gamma$ & $A$ & $B$ & $C$ & $A$ & $B$ & $C$ \\
\hline
SCDM & $\gamma=0$ & $0.56$ & $ 3.0$ & $2.5$ & $0.58$ & $ 2.9$ & $2.4$\\
{}   & $\gamma=1$ & $0.62$ & $ 2.7$ & $2.2$ & $0.64$ & $ 2.6$ & $2.1$\\
{}   & $\gamma=2$ & $0.70$ & $ 2.4$ & $1.9$ & $0.72$ & $ 2.3$ & $1.8$\\
OCDM & $\gamma=0$ & $0.59$ & $ 2.8$ & $2.3$ & $0.43$ & $ 2.1$ & $1.8$\\
{}   & $\gamma=1$ & $0.66$ & $ 2.5$ & $2.0$ & $0.72$ & $ 2.3$ & $1.8$\\
{}   & $\gamma=2$ & $0.50$ & $ 1.7$ & $1.5$ & $0.80$ & $ 2.1$ & $1.6$\\
$\Lambda$CDM & $\gamma=0$ & $0.65$ & $ 2.6$ & $2.1$ & $0.67$ & $ 2.5$ & $2.0$\\
{}   & $\gamma=1$ & $0.72$ & $ 2.4$ & $1.8$ & $0.75$ & $ 2.2$ & $1.7$\\
{}   & $\gamma=2$ & $0.81$ & $ 1.8$ & $1.6$ & $0.84$ & $ 1.9$ & $1.4$\\
\hline
\end{tabular}
\end{center}
\end{table}

This suggests that for a choice of the cosmological model and
$\gamma$, there exists a simple phenomenological law that relates $R$
to $\langle z\rangle$ and $\Delta z$, which is valid for all source
models considered here,
\begin{eqnarray}
\label{fp}
R=A\ {\Delta z^C\over \langle z\rangle^B},
\end{eqnarray}
where $A$ is of the order of $0.5$, $B$ and $C$ varying from 1.5 to 3.
The precise values of parameters $(A,B,C)$ are obtained by a least-squares
fitting method. They  are given in Table \ref{table:fp}.

\begin{figure}
\begin{center}
\begin{minipage}{8cm}
\epsfxsize=8cm \epsffile{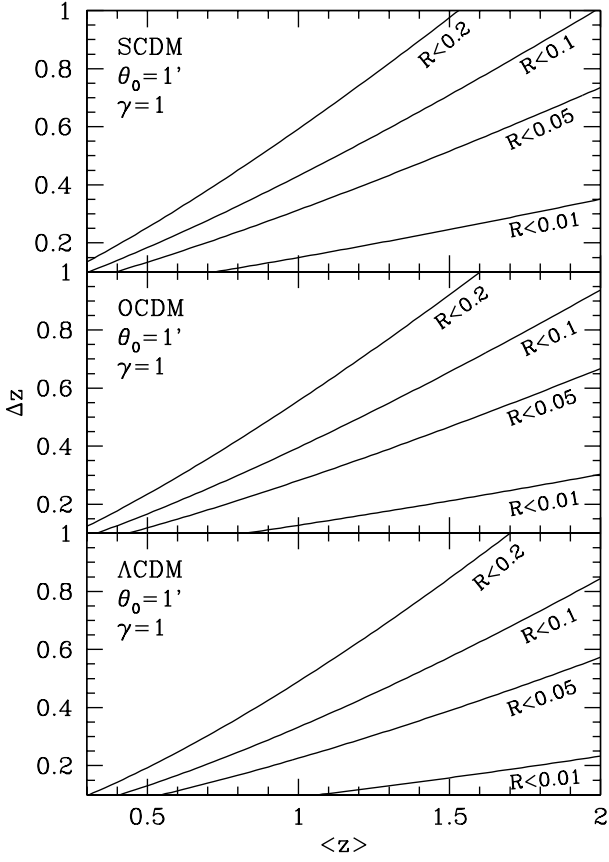}
\end{minipage}
\end{center}
\caption{Contour lines of $R$ obtained from our semi-analytic modeling
in $\langle z \rangle$-$\Delta z$ plane for $\theta=1'$ and
$\gamma=1$. Each panel corresponds to a different cosmology.  }
\label{fig:az_dz}
\end{figure}

The accuracy of this law is demonstrated in Figure  \ref{fig:rat_fp}.
One can see that the data lie fairly well on the parameterized
line. It allows us to make a contour plot of parameter $R$ in $\langle
z \rangle$-$\Delta z$ space, as shown in Fig.~\ref{fig:az_dz}, for
three cosmological models. Not surprisingly, this figure clearly
indicates that the way to reduce out SLC effect on a measurement of
the skewness is to make the source distribution narrow with a high
mean redshift.
 
\begin{figure}
\begin{center}
\begin{minipage}{8cm}
\epsfxsize=8cm \epsffile{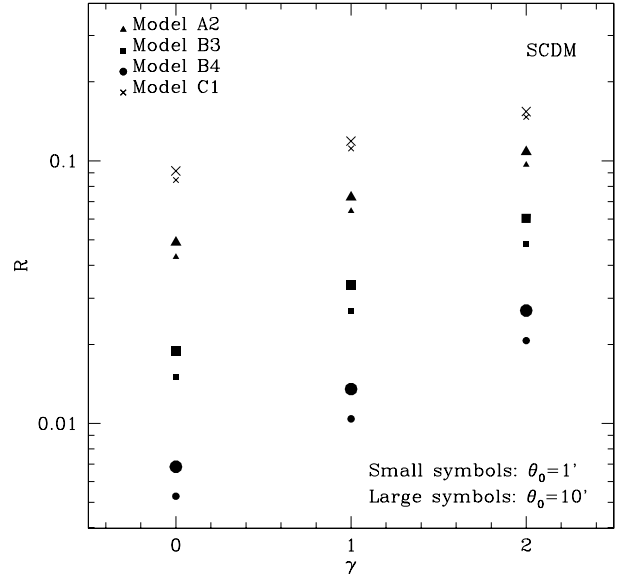}
\end{minipage}
\end{center}
\caption{Semi-analytic values of $R$ as a function of $\gamma$ for 4
selected models in SCDM.}
\label{fig:rat_b.s}
\end{figure}

\begin{figure}
\begin{center}
\begin{minipage}{8cm}
\epsfxsize=8cm \epsffile{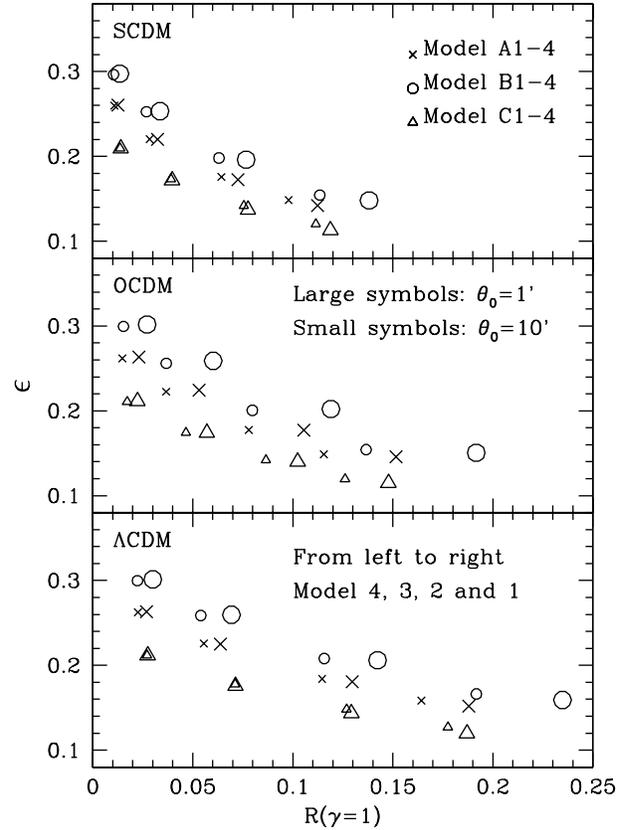}
\end{minipage}
\end{center}
\caption{The slopes of the $R$-$\gamma$ relation in the semi-analytic
model.  Coefficient $\epsilon$ as defined in equation (\ref{g-a}) is
shown  as a function of $R(\gamma=1)$.  }
\label{fig:a_rat}
\end{figure}

Finally, we examine the dependence of the SLC effect on the evolution of
bias.  Figure \ref{fig:rat_b.s} shows $R$ as a function of $\gamma$
for 4 models selected arbitrarily.  For each source distribution
model,  the $R$-$\gamma$ relation is  well fitted by the following
empirical law,
\begin{eqnarray}
\label{g-a}
\log R = \epsilon \gamma+\mbox{const}.
\end{eqnarray}
The coefficient $\epsilon$, which describes the strength of dependence
of the SLC effect on the evolution of bias, is shown in
Fig.~{\ref{fig:a_rat}} as a function of $R(\gamma=1$) for all source 
distribution models we consider.  
Each panel corresponds to a given choice of cosmology.  One
can see that $\epsilon$  remains in the range $0.1<\epsilon<0.3$ and
is almost insensitive to both smoothing scale and cosmology.  At
a fixed value of $R(\gamma=1)$, $\epsilon$ increases with the mean
redshift (in order of C, A and B).  
This is a natural consequence of the fact that, for our choice of
the bias evolution model eq.~(\ref{bias}), the
impact of the change in the bias evolution is more significant at
a higher redshift.

The uncertainty in parameter $R$ caused by our ignorance of   
$b(z)$ can be roughly estimated  using the empirical relation
(\ref{g-a}) as follows: suppose that the power-law model (\ref{bias})
for the evolution of bias stands, but that there is an error
$\Delta_{\gamma}$ on the value of $\gamma$.  Applying simple error
propagation technique, one finds $\delta R/R =
2.3\,\epsilon\,{\Delta_\gamma} (\sim0.5 \Delta_{\gamma})$: if one is
able to constrain the bias evolution model with an accuracy better
than $\Delta_{\gamma}<0.4$, the uncertainty in $R$ drops below 
20 per cent.

\section{Testing semi-analytic predictions against numerical 
simulations}
\label{sect:6}

In this section, we compare theoretical predictions to  ray-tracing
experiments in $N$-body simulations,  using mock galaxy catalogues
extracted from the simulations as distributions of sources.  The
$N$-body data sets and the ray-tracing method used for this work are
described in Appendix B. 
In Appendix C, details of the procedure to generate mock galaxy
catalogues are presented.  
We focus
only on the SCDM model, but the conclusions of our numerical analysis
should not depend significantly on the considered cosmology.

\begin{figure}
\begin{center}
\begin{minipage}{8cm}
\begin{center} 
\epsfxsize=8cm \epsffile{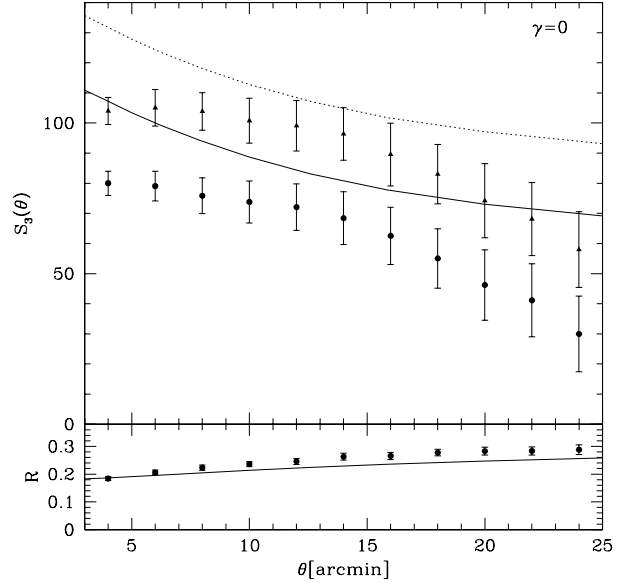}
\end{center}
\end{minipage}
\end{center}
\caption{$S_3$ (upper panel) and $R$ (lower panel)  as functions of
the smoothing angle $\theta$ in the $\gamma=0$ case.
The solid (dotted) line shows the semi-analytic prediction with (without) 
the SLC effect taking into account, while the filled circles (triangles)
show the results of  ray-tracing simulations with (without) SLC effect
taking into account.
Error bars give some estimate of the uncertainty $E$ on the measurements,  
i.e. $E(S_3) \simeq \Delta
S_3/\sqrt{40}$, where $(\Delta S_3)^2$  is the dispersion over the 40
realizations. Note that $R$ is calculated as
$R=-S_3^{\rm slc}/S_3^{\rm n.l.}$, where $S_3^{\rm n.l.}$ is always given by
semi-analytic predictions.}
\label{fig:mock_pre}
\end{figure}
We measure convergence statistics  on mock galaxy
catalogues as follows: the value of convergence for a galaxy at redshift
$z$  is given by linear interpolation
between the $\kappa_i$ ---computed from rays propagating between
redshift $z$ of the closest lens plane to the galaxy (see
Appendix B) and present time---   measured at the four
nearest angular pixels from the galaxy position in the sky.  
  The amplitude of the SLC effect is measured by comparing  
  simulations to  similar SLC-free ones. They are obtained from other
mock galaxy catalogues with  the same source distribution $n_s(z)$
(i.e. the same as in the SLC mock catalogue),  in which galaxies are
 randomly distributed on the sky.   Note finally that top hat filtering is
used  and the intrinsic ellipticity of galaxies is not taken into account.

The upper panel of Fig.~\ref{fig:mock_pre} shows the skewness parameter
measured from the $\gamma=0$ mock galaxy catalogue with and without the SLC
effect. The lower panel displays the function $R(\theta)$,
except that in the denominator of equation (\ref{R}), we always take
the value given by nonlinear semi-analytic predictions, $S_3^{\rm n.l.}$.
As discussed in Appendix B (see also Hamana \& Mellier 2001), the
simulations have limited available dynamic range since they are
contaminated by force softening and finite volume effects 
at small and large scales respectively, where the measured $S_3$ is expected to
underestimate the real value. 
Furthermore, there is a 10-20 per cent uncertainty in the nonlinear
perturbation theory  predictions. With these elements in mind, we see
that agreement between measurements and predictions is reasonable
when  SLC effects are taken into account.  In particular, the order of
magnitude of the shift between the upper and the lower symbols in the top
panel of Fig.~\ref{fig:mock_pre} matches very well that between the
dotted and the solid curve, as illustrated by bottom panel.

Similar results are obtained for the $\gamma=1$ and $2$ catalogues, as
summarized in Fig.~\ref{fig:mock_deltas3}, which concentrates on the
parameter $R$. 
Since numerical experiments for $\gamma=1$ and 2 cases are done with
threshold bias instead of the linear bias used in the semi-analytic
computation, we should only focus on the differences among the three models
arising from different evolution of biasing.
Although there is a systematic difference between the predictions and
measurements, a trend in the dependence of
evolution of biasing on the SLC effect found in the predictions is
well reproduced in the measurements.

Furthermore, one finds that the $\gamma=2$ measurement agrees
better with the $\gamma=3$ prediction except for the largest angular scales.
This is consistent with the scale dependence of biasing detected in
the $\gamma=2$ mock catalogue (on small scales the bias evolves as $\gamma=3$,
see Appendix C for detailed discussion on this point). 
We may conclude from the results above that the semi-analytic approach
gives a good prediction of the SLC effect on the convergence
skewness. 

\begin{figure}
\begin{center}
\begin{minipage}{8cm}
\begin{center} 
\epsfxsize=8cm \epsffile{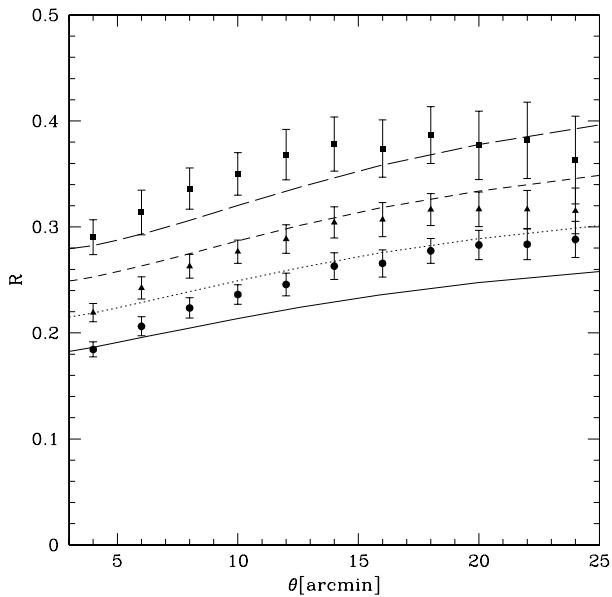}
\end{center}
\end{minipage}
\end{center}
\caption{The parameter $R$ as displayed in the lower panel of
Fig.~\ref{fig:mock_pre} but now for all values of $\gamma$
considered. Symbols show the results of ray-tracing simulations:
circles, triangles and squares correspond respectively to $\gamma=0$,
$1$ and $2$ mock catalogues. Lines correspond to semi-analytical
predictions, $\gamma=0$, 1, 2, 3 from bottom to top. Error bars are
computed as explained in caption of  Fig.~\ref{fig:mock_pre}.}
\label{fig:mock_deltas3}
\end{figure}

\section{Summary and discussion}

We have examined the source-lens clustering (SLC) effect on measurements of the
skewness of lensing convergence using a nonlinear semi-analytic
approach.  The result of semi-analytic predictions were tested against
numerical  simulations, and a good agreement between them was found.
Our main conclusions are as follows:
\begin{itemize}
\item SLC effect strongly depends on the redshift  distribution of
source galaxies.  We found that the effect scales with the width and
mean redshift of the  distribution roughly as  $R \propto \langle z
\rangle^{-(3.0-1.8)} \Delta z^{1.4-2.5}$, (where $R=-S_3^{\rm slc}/S_3$,
and $S_3^{\rm slc}$ is the change in measured $S_3$ due to SLC).  As
illustrated by Fig.~\ref{fig:az_dz}, this relation indicates that it
is  essential to make the width of the distribution narrow and 
its mean redshift high to reduce the SLC effect (this was partly
pointed out by  B98).
\item SLC effect also depends on the evolution of the bias between the
galaxy and total matter distributions, $b(z)$.  Assuming a simple
power-law model and linear bias, $b(z) \propto (1+z)^{\gamma}$, we
found that the uncertainty in  $\gamma$ transforms into $\delta R/R =
2.3\,\epsilon\, \Delta_{\gamma}$  with a typical value of $\epsilon
\sim0.2$.  
This indicates that the uncertainty in $\gamma$ must be
$\Delta_{\gamma}<0.2$ for predicting the amplitude of the SLC  effect
with better than 10 per cent accuracy.
\end{itemize}

The main uncertainty in  semi-analytic predictions comes from the fact
that the accuracy of the nonlinear fitting formula of the density
bispectrum is only 10-20 per cent. We  expect the same level of uncertainty in
predictions presented  in this paper for the  SLC effect on the
convergence skewness.  
This can actually be improved by measurements  in large
$N$-body simulations with high spatial resolution.

Since, so far, little is known about the evolution of the bias, it is
still very difficult to predict the SLC effect accurately.  It is therefore
very important to reduce this effect as much as possible by
controlling the redshift distribution of sources.  
The above results tells us that an ideal observational strategy  might
be as follows:  (i) going to a deep limiting magnitude to increase the
mean
redshift of the survey and  (ii) using only fainter images to reduce the
width of the distribution.  A desirable source distribution for
$R<0.1$ suggested by Fig.~\ref{fig:az_dz} would have  $\Delta z<0.3$
and $\langle z \rangle>1$.  This may be of course challenging: going
to deeper magnitudes will make  the calculation of the redshift
distribution of sources more difficult, and using only faint images
will increase the noise due to intrinsic ellipticity of galaxies.
We leave more detailed studies on the designing of optimal strategies to
future works.

Before closing this paper, it is important to emphasize that the use
of source galaxies with a small redshift width may unfortunately introduce 
additional skewness signal due to the intrinsic correlation of galaxy
ellipticities.
It was indeed suggested that the amplitude of this latter effect 
scales roughly as $(\Delta z)^{-1}$, although the
normalization of this relation is ambiguous because of the uncertainty
in the correlation between the shape of galaxies and that  
of their dark matter halos (Croft \& Metzler 2000; 
Crittenden et al.~2000; Heavens, Refregier \&
Heymans 2000; Pen, Lee \& Seljak 2000; Catelan, Kamiokowski \&
Blandford 2001; Hatton \& Ninin 2001).
However, as pointed out by Croft \& Metzler (2000), 
intrinsic ellipticity correlations might just
act as an additional source of random noise, without
significantly influencing the measured value of the skewness 
of the convergence. 

\section*{Acknowledgments}
We would like to thank L.~Van Waerbeke for providing the FORTRAN code
to compute the nonlinear skewness of convergence and for helpful
comments. We would also like to thank A. Stebbins for teaching us his
way of generating mock galaxy catalogues from $N$-body simulations and
S.~Hatton for very useful suggestions to improve the text.
This research was supported in part by the Direction de la Recherche
du Minist{\`e}re Fran{\c c}ais de la Recherche.   The computational
means (CRAY-98 and NEC-SX5) to do the $N$-body simulations  were made available
to us thanks to the scientific council of  the Institut du
D\'eveloppement et des Ressources en Informatique  Scientifique
(IDRIS).  Numerical computation in this work was partly carried out at
IAP at the TERAPIX data center and on MAGIQUE (SGI-02K).


\appendix

\section{Perturbation theory approach to the cosmic shear statistics
in the presence of SLC}
\label{sect:Appendix A}

The expressions for the skewness of lensing convergence and the
correlation term due to SLC were first derived by BvWM97 and B98,
respectively, in the framework of perturbation theory.
However, B98 only gave the expression for the case of an Einstein-de
Sitter cosmological model and assumed power-law density power spectrum.
In this Appendix, for this paper being self-contained, we re-derive
the skewness terms which are valid for an arbitrary Friedmann model.
We basically follow the original deviation by B98.
It should be noted that the skewness corrections are not only caused
by the SLC but also arise from e.g., the lens-lens coupling (BvWM97,
Van Waerbeke et al 2001b) and the lensing magnification effect (Hamana
2001). 
In what follows, we focus on the SLC correction and are not concerned
with other correction terms.

\subsection{Fluctuation in a source distribution due to the source
clustering} 

The number density of the sources at redshift $z$ and in 
direction $\bmath{\phi}$ can be written, 
\begin{eqnarray}
\label{eq:ap:ns}
n_s^{\rm obs}[\chi(z),\bmath{\phi}] =
n_s[\chi(z)]\left(1+\delta_s[\chi(z),\bmath{\phi}]\right),
\end{eqnarray}
where $n_s[\chi(z)]$ is the average number density of sources,
$\delta_s[\chi(z),\bmath{\phi}]$ is their local density contrast and
$\chi$ denotes the radial comoving distance.
We suppose as B98 that the average source number density is normalized to
unity, $\int_0^{\chi_H} d\chi n_s(\chi)=1$, where $\chi_H$ is the
distance to the horizon (the normalized distribution denotes the
probability distribution).
Following B98, we assume that the density contrast of sources
is related to the matter density contrast, $\delta$, via the linear biasing,
\begin{eqnarray}
\label{eq:ap:bias}
\delta_s(\chi,\bmath{\phi}) = 
b(\chi)\delta(\chi,\bmath{\phi}).
\end{eqnarray}

\subsection{Convergence statistics in the presence of SLC}

Let us consider the measured convergence that results from averages
made over many distant galaxies located at different distances.
Denoting the smoothing scale by $\theta$, such an average can formally
be written as,
\begin{eqnarray}
\label{average-kappa}
\kappa_\theta={
{\sum_{i=1}^{N_s} W_\theta(\bmath{\phi}_i)
\kappa_s(z_i,\bmath{\phi}_i)}
\over
{\sum_{i=1}^{N_s} W_\theta(\bmath{\phi}_i)}
},
\end{eqnarray}
where $W_\theta(x)$ denotes the weight function of the average,
$N_s$ is the number of source galaxies, $\kappa_s(z_i,\bmath{\phi}_i)$ is
the lensing convergence signal from a source galaxy located at
redshift $z_i$ in a direction $\bmath{\phi}_i$ and is given by
(e.g., Mellier 1999; Bartelmenn \& Schneider 2001 for reviews)
\begin{eqnarray}
\label{eq:ap:kappa_s}
\kappa_s(z,\bmath{\phi}) =
{{3 \Omega_{\rm m}} \over 2}{{H_0} \over c}
\int_0^{\chi_s(z)} d\chi_l\, g(\chi_l,\chi_s) \delta(\chi_l,\bmath{\phi}),
\end{eqnarray}
with
\begin{eqnarray}
\label{eq:ap:g}
g(\chi_l,\chi_s) = {{H_0}\over c} {{f(\chi_l) f(\chi_s-\chi_l)} \over 
{f(\chi_s) a(\chi_l)}}.
\end{eqnarray}
Here $a$ is the scale
factor normalized to its present value, and $f(\chi)$ denotes
the comoving angular diameter distance,
defined as $f(\chi)=K^{-1/2} \sin K^{1/2} \chi$, $\chi$, $(-K)^{-1/2}
\sinh (-K)^{1/2} \chi$ for $K>0$, $K=0$, $K<0$, respectively, where
$K$ is the curvature which can be expressed as
$K=(H_0/c)^2(\Omega_{\rm m}+\Omega_\lambda-1)$.
For the weight function, the angular top-hat filter (BvWM97) and/or
compensated filter (Schneider et al.~1998) are commonly adopted (e.g.,
Van Waerbeke et al.~2001a).
In what follows, we consider the top-hat filter for the weight
function, and in this case equation (\ref{average-kappa}) is reduced to
$\kappa_\theta=\sum_{i=1}^{N_s^j} \kappa_s(z_i,\bmath{\phi}_i)/N_s^j$,
where $N_s^j$ is the number of source galaxies within an aperture $\theta$
centered on a direction $\bmath{\phi}_j$.
The number density of sources for current and future weak lensing
analyses is about 40 per  arcmin$^2$ (e.g., Van Waerbeke et
al.~2001a), which typically implies more than 100 galaxies 
in discs of radius $\theta \geq  1$ arcmin. As a result, discreteness
effects from the source distribution can be neglected (see also
B98) and we can rewrite (\ref{average-kappa}) in the continuous limit:
\begin{eqnarray}
\label{average-kappa2}
\kappa_\theta={
{\int d^2 \phi\, W_\theta(\bmath{\phi}) \int_0^{\chi_H} d \chi\, 
\kappa_s(\chi,\bmath{\phi}) n_s^{\rm obs}(\chi,\bmath{\phi})}
\over
{\int d^2 \phi\, W_\theta(\bmath{\phi}) \int_0^{\chi_H} d \chi\,
n_s^{\rm obs}(\chi,\bmath{\phi})}
}.
\end{eqnarray}
Let us now expand equation (\ref{average-kappa2}) in terms of
$\delta$ using the perturbation theory approach, following BvWM97.
The presence of SLC does not change the
expression of the first order term,
\begin{eqnarray}
\label{k1}
\kappa_\theta^{(1)} &=& {{3 \Omega_{\rm m}} \over 2} {{H_0} \over c}
\int d^2 \phi\, W_\theta(\bmath{\phi})
\int_0^{\chi_H} d \chi_s\, {n_s(\chi_s)}\nonumber\\
&&\times\int_0^{\chi_s} d \chi_l\,
g(\chi_l,\chi_s) \delta^{(1)}(\chi_l,\bmath{\phi})\nonumber\\
&=& {{H_0} \over c} \int d^2 \phi\, W_\theta(\bmath{\phi})
\int_0^{\chi_H} d \chi_l\,w(\chi_l)\delta^{(1)}(\chi_l,\bmath{\phi}),
\end{eqnarray}
where $w(\chi)$ is so-called the lensing efficiency function defined by
\begin{eqnarray}
\label{effeciency}
w(\chi_l) = {{3 \Omega_{\rm m}} \over 2} 
\int_{\chi_l}^{\chi_H} d\chi_s
g(\chi_l,\chi_s)
{n_s(\chi_s)}.
\end{eqnarray}
The second order convergence consists of two terms:
one comes from the second order density perturbation and it is
formally written by replacing the subscript $^{(1)}$ in the first order
expression (\ref{k1}) with $^{(2)}$ (BvWM97);
the other one is due to SLC, 
\begin{eqnarray}
\label{kmag}
\kappa_\theta^{\rm slc(2)}
&=& {{3 \Omega_{\rm m}} \over 2} {{H_0} \over c}
\int d^2 \phi\, W_\theta(\bmath{\phi})\nonumber\\
&&\times \int_0^{\chi_H} d \chi\, {n_s(\chi)}
b(\chi)\delta^{(1)}(\chi,\bmath{\phi})\nonumber\\
&&\times \int_0^{\chi} d \chi'\,
g(\chi',\chi) \delta^{(1)}(\chi',\bmath{\phi})\nonumber\\
&&-\kappa_\theta^{(1)} 
\int d^2 \phi\, W_\theta(\bmath{\phi})\nonumber\\
&&\times \int_0^{\chi_H} d \chi\, {n_s(\chi)}
b(\chi)\delta^{(1)}(\chi,\bmath{\phi}).
\end{eqnarray}
Using the small angle approximation (Kaiser 1992),
equation (\ref{k1}) is rewritten in terms of the Fourier transform of
the density contrast, $\delta(k)$, as
\begin{eqnarray}
\label{k1f}
\kappa_\theta^{(1)} &=& {{H_0} \over c} \int_0^{\chi_H} d
\chi\,w(\chi)
\int {{d^3 k} \over {(2\pi)^3}} \delta^{(1)}[\bmath{k};\chi] 
\exp[i k_{\chi} f(\chi)]\nonumber\\
&& \times W[f (\chi) k_\perp \theta],
\end{eqnarray}
where the wave vector $\bmath{k}$ is decomposed into the line-of-sight
component $k_\chi$ and its perpendicular, $\bmath{k_\perp}$, and
$W(x)$ is the Fourier transform of the weight function.
In the case of the top-hat filter, $W(x)=2J_1(x)/x$ where $J_1$ is
the Bessel function of first order. 
In the same manner, equation (\ref{kmag}) reads
\begin{eqnarray}
\label{kmagf}
\kappa_\theta^{\rm slc(2)}
&=& {{3 \Omega_{\rm m}} \over 2} {{H_0} \over c}
\int_0^{\chi_H} d \chi\, {n_s(\chi)}b(\chi)
\int_0^{\chi} d \chi'\, g(\chi',\chi) \nonumber\\
&&\times \int {{d^3 k} \over {(2\pi)^3}} \delta^{(1)}[\bmath{k};\chi] 
\exp[i {k}_{\chi} f(\chi)]  \nonumber\\
&&\times\int {{d^3 k'} \over {(2\pi)^3}} \delta^{(1)}[\bmath{k'};\chi'] 
\exp[i {k'}_{\chi}f(\chi')] \nonumber\\
&&\times W[|f (\chi) \bmath{{k}_\perp} + f (\chi') \bmath{{k'}_\perp}|\theta]
\nonumber\\
&&-\kappa_\theta^{(1)} 
\int_0^{\chi_H} d \chi\, {n_s(\chi)} b(\chi) \nonumber\\
&& \times \int {{d^3 k} \over {(2\pi)^3}} \delta^{(1)}[\bmath{k};\chi] 
\exp[i {k}_{\chi}
f(\chi)] W[f (\chi) {k}_\perp \theta].\nonumber\\ 
\end{eqnarray}

The average of the convergence is not affected by
the presence of the SLC and is therefore zero, $\langle
\kappa \rangle =0$. 
The variance is not affected by it either at linear order and is given by
\begin{eqnarray}
\label{variance}
V_\kappa (\theta)
&=&\langle {\kappa_\theta^{(1)}}^2 \rangle\nonumber\\
&=& \left({{H_0}\over c}\right)^2  \int_0^{\chi_{H}} 
d\chi_l\,w^2(\chi_l) I_0(\chi_l,\theta),
\end{eqnarray}
with
\begin{eqnarray}
\label{eq:ap:I0}
I_0(\chi,\theta) = {1\over {2\pi}} \int dk\, k P_{\rm lin} (\chi,k)  
W_{\rm 2D}^2 [k f(\chi) \theta],
\end{eqnarray}
where $P_{\rm lin} [\chi(z),k]$ is the linear density power spectrum.  
In the presence of SLC, the skewness parameter,
defined by $S_3(\theta)=\langle \kappa_\theta^3 \rangle/V_\kappa^2(\theta)$,
consists of two terms:
one comes from the second order perturbation (BvWM97),
\begin{eqnarray}
\label{eq:ap:kap3_2pt}
\langle \kappa_\theta^3\rangle^{q.l.} &=&
3 \langle {\kappa_\theta^{(1)}}^2 \kappa_\theta^{(2)}\rangle\\
&=& 6 \left({{H_0}\over c}\right)^3 \int_0^{\chi_{H}} d
\chi\, w^3(\chi) \nonumber\\ && \times \left[ {6\over7}
I_0^2(\chi,\theta)+{1\over4} I_0(\chi,\theta)
I_1(\chi,\theta)\right],
\end{eqnarray}
where
\begin{eqnarray}
\label{I1}
I_1(\chi,\theta) = {1\over {2\pi}}
\int dk\, k^2 P_{\rm lin} (\chi,k) {{dW_{\rm 2D}^2 [k f(\chi) \theta]} 
\over {dk}}.
\end{eqnarray}
The other arises from SLC, 
\begin{eqnarray}
\label{s3mag}
\langle \kappa_\theta^3\rangle^{\rm slc}&=&3\langle
{\kappa_\theta^{(1)}}^2 \kappa_\theta^{\rm slc(2)} \rangle\nonumber\\
&=& 9 \Omega_{\rm m} \left({{H_0}\over c}\right)^3
\int_0^{\chi_H} d \chi\, {n_s(\chi)}  b(\chi)
w(\chi) I_0(\chi) \nonumber\\
&& \times \int_0^{\chi} d \chi'\, g(\chi',\chi) w(\chi') 
I_0(\chi')\nonumber\\
&&-6 V_\kappa(\theta) {{H_0} \over c}
\int_0^{\chi_H} d \chi\, {n_s(\chi)} b(\chi)
w(\chi) I_0(\chi).\nonumber\\
\end{eqnarray}
To derive the last expression, we used an approximation, which turns
out to be very accurate for top-hat smoothing (see B98 for details),
\begin{eqnarray}
\label{eq:appro}
{1 \over {2\pi}} \int_0^{2\pi}d\vartheta \sin\vartheta
W(|\bmath{k}+\bmath{k'}|)\simeq W(k)W(k'),
\end{eqnarray}
where $\vartheta$ is
the angle between the wave vectors $\bmath{k}$ and $\bmath{k'}$ and
$k=\vert \bmath{k} \vert$. 

Then, the convergence skewness simply reads, in the second order
perturbation theory framework,  $S_3(\theta)=S_3^{\rm q.l.}+S_3^{\rm slc}=\langle
\kappa_\theta^3\rangle^{\rm q.l.}/V^2_\kappa(\theta)+\langle
\kappa_\theta^3\rangle^{\rm slc}/V^2_\kappa(\theta)$, where all the
terms are computed above.

Note that our calculation is slightly different from that of B98. Indeed B98
assumed an optimally weighted estimator for the convergence leading to
eq.~(9) of his paper instead of our eq.~(\ref{average-kappa2}). 
However, with approximation
(\ref{eq:appro}), both estimators give the same results:
eq.~(\ref{kmagf}) would match eq.~(22) of B98, and
therefore we would easily recover eq.~(29) of B98 for a scale-free 
power-spectrum\footnote{Notice the difference of sign convention we
use for $\kappa$, to enforce positively for the convergence skewness.}.

\section{A brief description of the ray-tracing simulations}
\label{sect:Appendix B}

In this Appendix, we describe the $N$-body data sets and the
ray-tracing method used for this work. More technical details  
are presented in Hamana \& Mellier (2001).

Light ray trajectories are followed through  large $N$-body
simulations data set generated with a  fully vectorized and
parallelized Particle-Mesh (PM) code.
Each $N$-body experiment involves
$256^2\times512$ particles in a periodic rectangular box of size
$(L,L,2L)$.  The mesh used to compute the forces was   $256^2\times
512$. A light-cone of the particles was extracted from each simulation
during the run as explained in Hamana, Colombi \& Suto (2001).
Our aim was for the
light-cone to cover a large redshift range, $0 \leq z \la 3$, and a
field of view of $5\times5$  square degrees.  To do that, we adopted
the {\it tiling}  technique first proposed by White \& Hu (2000): we
performed 11 independent simulations  covering adjacent redshift
intervals $[z_i^{\rm min}, z_i^{\rm max}]$, $i=1,\ldots,11$. The
size of each simulation is such that  the portion of the light-cone in
$[z_i^{\rm min}, z_i^{\rm max}]$ (aligned with the third axis) exactly
 fits the box-size. This way, angular resolution is
approximately conserved as a function of redshift, except close to the
observer.  Finally, in order to have enough structures in each box,  we impose
the supplementary constraint $L\geq 80h^{-1}$ Mpc. As a result, $L$
follows the following sequence with redshift,
$80,80,80,80,80,120,160,240,320,480,640\ h^{-1}$ Mpc.

The multiple lens-plane algorithm was used for ray-tracing
calculations (Jain et al.~2000 and references therein).  The lens
planes (which are, at the same time, source planes) are  separated by
intervals of $80h^{-1}$Mpc, amounting to a total number of 38 in the
redshift range $0 \leq z \la 3$.  For each ray, the lensing
magnification matrix is computed on the source planes and is stored.
We performed 40 realizations of the underlying density field  by
random shifts of the simulation boxes in the $(x,y)$ plane.  For each
realization, $512^2$ rays are traced backward from the observer.  The
initial ray directions are set on $512^2$ grids, which correspond to
pixels of angular size $5^\circ/512\sim 0.59$ arcmin.

\begin{figure}
\begin{center}
\begin{minipage}{8cm}
\begin{center} 
\epsfxsize=8cm \epsffile{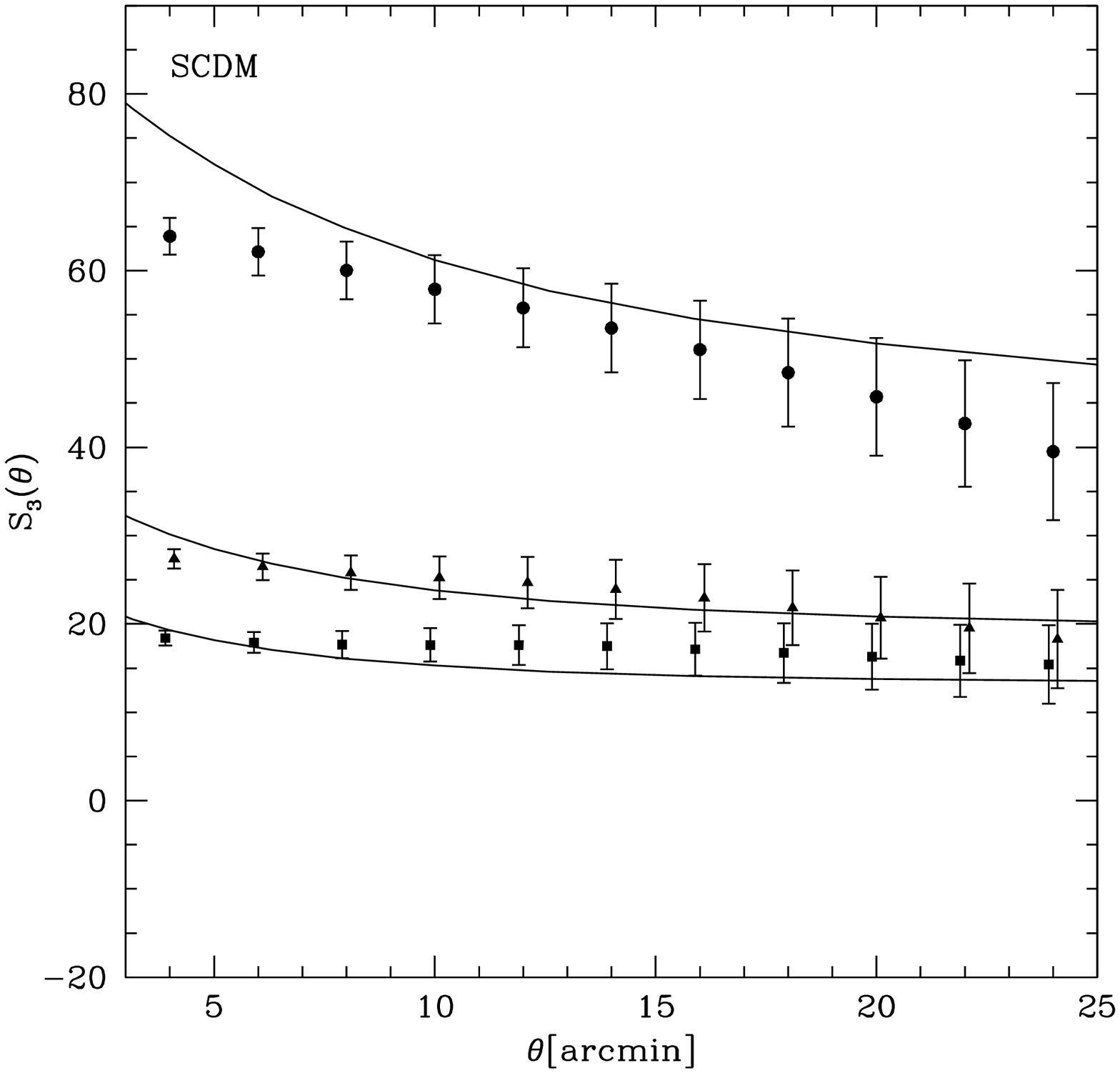}
\end{center}
\end{minipage}
\end{center}
\caption{The skewness $S_3$ of the lensing convergence measured from
the ray-tracing simulations (symbols)  compared with nonlinear
predictions (solid lines) for single source redshifts  $z_s\sim1$
(top), $2$, and $3$ (bottom).}
\label{fig:skew_zs}
\end{figure}

Before using realistic redshift distribution of sources, we compute
the skewness of the lensing convergence for single  source planes,
i.e., $n_s(z)=\delta_D(z=z_s)$ where $\delta_D$ is the Dirac delta
function. At this stage, we do not take into account the SLC effect.  
Figure
\ref{fig:skew_zs} shows $S_3$ obtained from the simulations compared
to nonlinear predictions.  Measurements match theory reasonably well,
as expected (Van Waerbeke et al.~2001b).  There are slight differences
which can be explained as follows:
\begin{enumerate}
\item The $N$-body simulations have a finite spatial resolution, which
implies a flattening of $S_3$ at scales smaller than about 4 arcmin.
\item At large angular scales, $\theta \ga 20-40$ arcmin, depending on
the source redshift considered (Hamana \& Mellier 2001), the measured
$S_3$ underestimates the real value, due to finite volume effects
(i.e. the lack of the large scale power, which contributes to the
skewness on smaller scales, due to the finite size of the simulation
boxes, e.g., Colombi, Bouchet \& Schaeffer 1994; see also Seto 1999).
\item There is an uncertainty in the fitting formula of the density
bispectrum (section 3), which transforms into a 10-20 per cent  error
on the semi-analytic prediction for $S_3$.  The differences between
theory and measurements in  Fig.~\ref{fig:skew_zs} are smaller than
this expectation, at least in the range where the measurements are
reliable, $4 \la \theta \la 20-40$ arcmin 
(derived from the above discussion on spatial
resolution and finite volume effects).
It is important to note this range is not equal to the dynamical
range that the ray-tracing simulation originally have, which is much
wider (see Hamana \& Mellier 2001 for discussion on this point).
\end{enumerate}
In conclusion, without SLC effects (yet) taken into account, the
semi-analytic prediction obtained for $S_3$ is accurate (Van Waerbeke
et al.~2001b).

\section{Procedure to generate mock galaxy catalogues}
\label{sect:Appendix C}

We generated three mock galaxy catalogues with galaxy number
counts $n_{\rm s}(z)$ derived from the semi-analytic model by
Devriendt \& Guiderdoni (2000) and reproducing as well as possible the
power-law model for the function $b(z)$, $b(z)=(1+z)^\gamma$ with
$\gamma=0$, 1, and 2. They are extracted from exactly the same dark
matter  distributions as the ones used for the ray-tracing simulations.

The procedure to create the mock catalogues can be described as follows:
\begin{enumerate}
\item We adopt threshold biasing, i.e. for a smooth density
distribution of dark matter, we assume that galaxies lie in regions of
density contrast larger than some threshold which may eventually depend on
redshift [point (ii) below],  $\delta \geq \delta_{\rm TH}(z)$. Inside
 these regions, the local number density of galaxies at
position $(z,\theta,\phi)$, (where $z$ denotes the redshift and
$(\theta,\phi)$ denote the direction in the sky) 
is proportional to dark matter density
\begin{equation}
   n_{\rm g}(z,\theta,\phi)=\mu(z)[1+\delta(z,\theta,\phi)].
  \label{eq:locbiasing}
\end{equation}
The normalization factor $\mu(z)$ is such that the redshift
distribution of galaxies reproduces (in terms of an ensemble average)
some prior, $n_{\rm s}(z)$, discussed in (iii).  To estimate the local
density contrast from our discrete dark-matter particle distribution,
we use local adaptive smoothing:  the mean quadratic distance $d$
between each simulation particle and its 6 nearest neighbors is
computed, $d^2=\sum_{i=1}^6 d_i^2$ where $d_i$ is the separation
between a central particle and $i$-the nearest neighbors. 
Then, $1+\delta \propto d^{-3}$.  For each dark matter
particle in regions with $\delta > \delta_{\rm TH}$,  $N$ 
galaxies are randomly placed in the sphere of radius $d$ centered on
the particle position.  $N$ is computed from a random realization of a
Poisson distribution with average ${\bar N}=(4/3)\pi d^3 n_{\rm g}$,
where $n_{\rm g}$ is the mean number density of galaxies.
\item Function $\delta_{\rm TH}(z)$ is determined numerically so that
the measured variances of density fluctuations in a sphere of radius
$8h^{-1}$ Mpc in the galaxy and the dark matter distribution,
respectively $\sigma_8^{\rm gal.}$ and $\sigma_8$, satisfy
\begin{equation}
\label{biasdef}
{{\sigma_8^{\rm gal.}(z)}\over{\sigma_8(z)}}=b(z).
\end{equation}
To do that, we use snapshots of the simulations at various redshifts
$z_i$, and compute iteratively $\delta_{\rm TH}(z_i)$ to match
equation (\ref{biasdef}) within a 3 per cent accuracy.  Then function
$\delta_{\rm TH}(z)$ is obtained by linear interpolation of the
$\delta_{\rm TH}(z_i)$. 
Note that since for $\gamma=0$, $b=1$ irrespective of the redshift,
the galaxy distribution directly traces the
matter distribution, and thus we do not need to have threshold i.e.,
we simply set $\delta_{\rm TH}=\delta_{\rm min}-1$. 
\item A prior function $n_s(z)$ is needed to compute the normalization factor
$\mu(z)$ equation (\ref{eq:locbiasing}). 
We have used the {\em ab-initio} semi-analytic approach to galaxy formation 
described in Devriendt \& Guiderdoni (2000) to obtain a reasonably 
realistic estimate of this function.
Such an approach is based on a Press--Schechter like
prescription to compute the number of galaxies as a function of
redshift, coupled to spectro-photometric evolution of stellar 
populations to calculate their luminosities. 
The results naturally match observed galaxy number counts and redshift 
distributions, as well as the diffuse extragalactic background light for
wavelengths ranging from the $UV$ to the near $IR$.
Here, we suppose that galaxies are selected in the
$I$ band, down to the magnitude  $I_{AB}=24.5$.  As a result, the
final mock catalogues yields a typical surface number density of 29
sources per arcmin$^2$  distributed in redshift as shown in  Figure
\ref{fig:mock_pz}.  The distribution has a peak at $z\sim 0.4$, with
mean redshift  $\langle z \rangle \sim 0.8$ and typical width $\Delta
z \sim 0.6$.
\end{enumerate}
Note that, for $b>1$, the linear bias prescription breaks down, because
$\delta_s$ can not be less than $-1$ by definition (in the
perturbation theory approach, this does not cause a serious problem
because $\vert \delta \vert $ is supposed to be much less than unity).
Therefore, we do not take the linear bias but use the threshold bias
for the cases of $\gamma=1$ and 2.
Accordingly, strictly speaking, the direct comparison between the
semi-analytical prediction and the numerical simulation makes sense
only for the $\gamma=0$ model.
However, we take $\gamma=1$ and 2 models to test the effect of the redshift
evolution of bias which should not depend strongly on details of the biasing
prescription. 
Moreover, to test the robustness of semi-analytic predictions in which
the simple linear bias is used, it is interesting to use a different biasing
prescription for the numerical experiments.

\begin{figure}
\begin{center}
\begin{minipage}{8cm}
\begin{center} 
\epsfxsize=8cm \epsffile{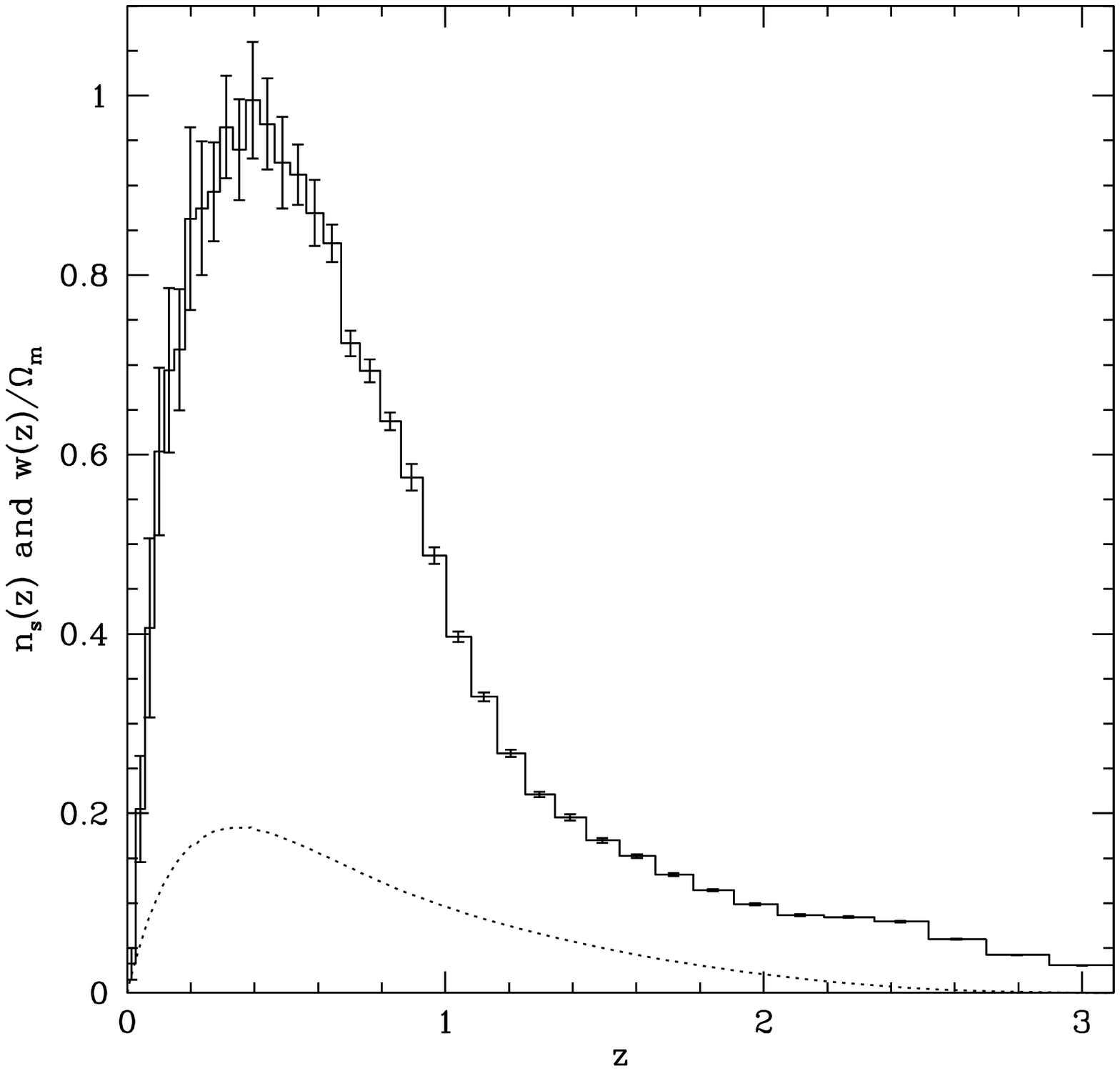}
\end{center}
\end{minipage}
\caption{The distribution of sources (histogram with error bars)  and
the lensing efficiency function (dotted line) as functions of
redshift. Error bars denote standard deviation computed among 40
realizations as discussed in Appendix B.  }
\label{fig:mock_pz}
\end{center}
\end{figure}

\begin{figure}
\begin{center}
\begin{minipage}{8cm}
\begin{center} 
\epsfxsize=8cm \epsffile{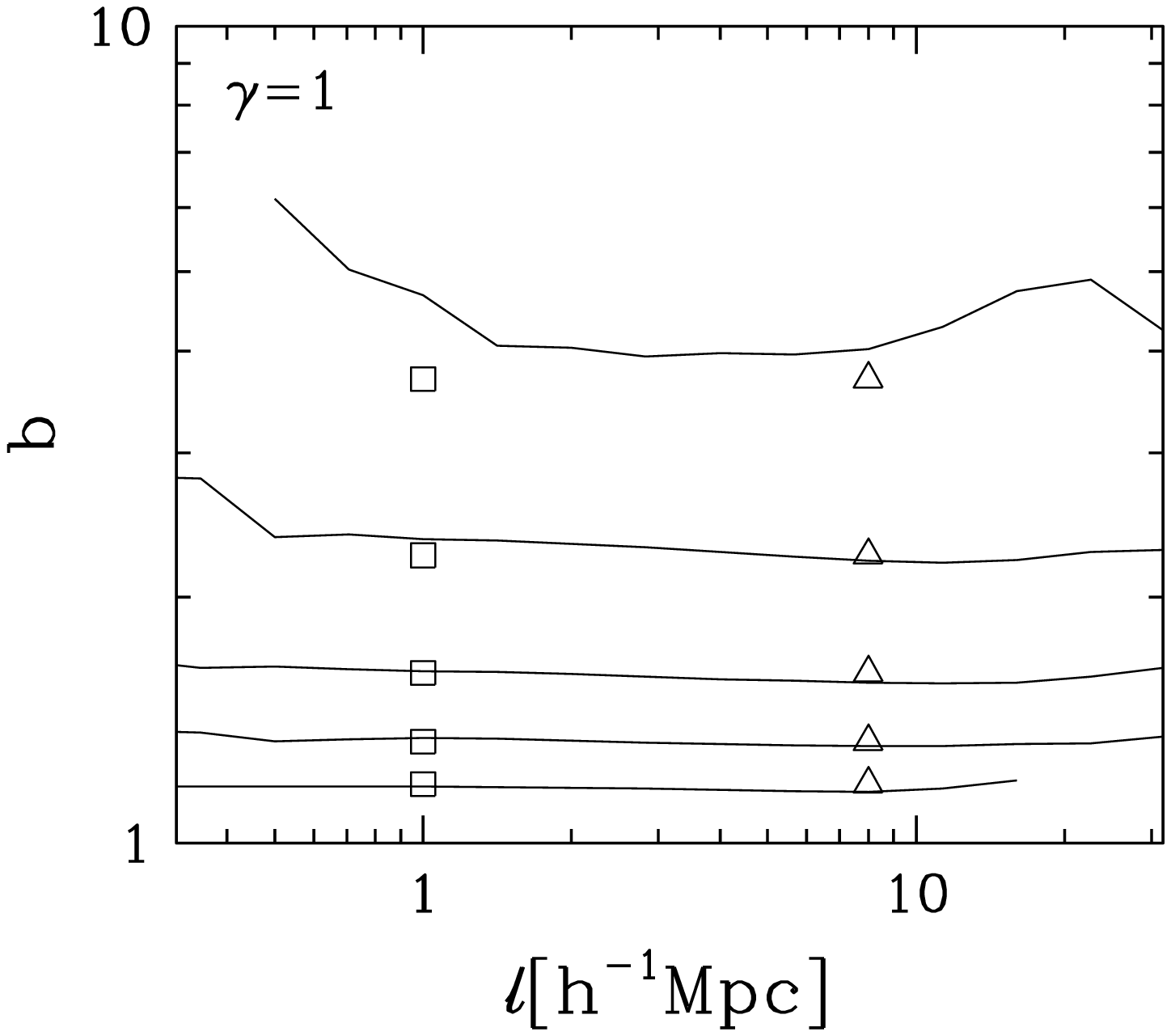}
\end{center}
\end{minipage}
\begin{minipage}{8cm}
\begin{center} 
\epsfxsize=8cm \epsffile{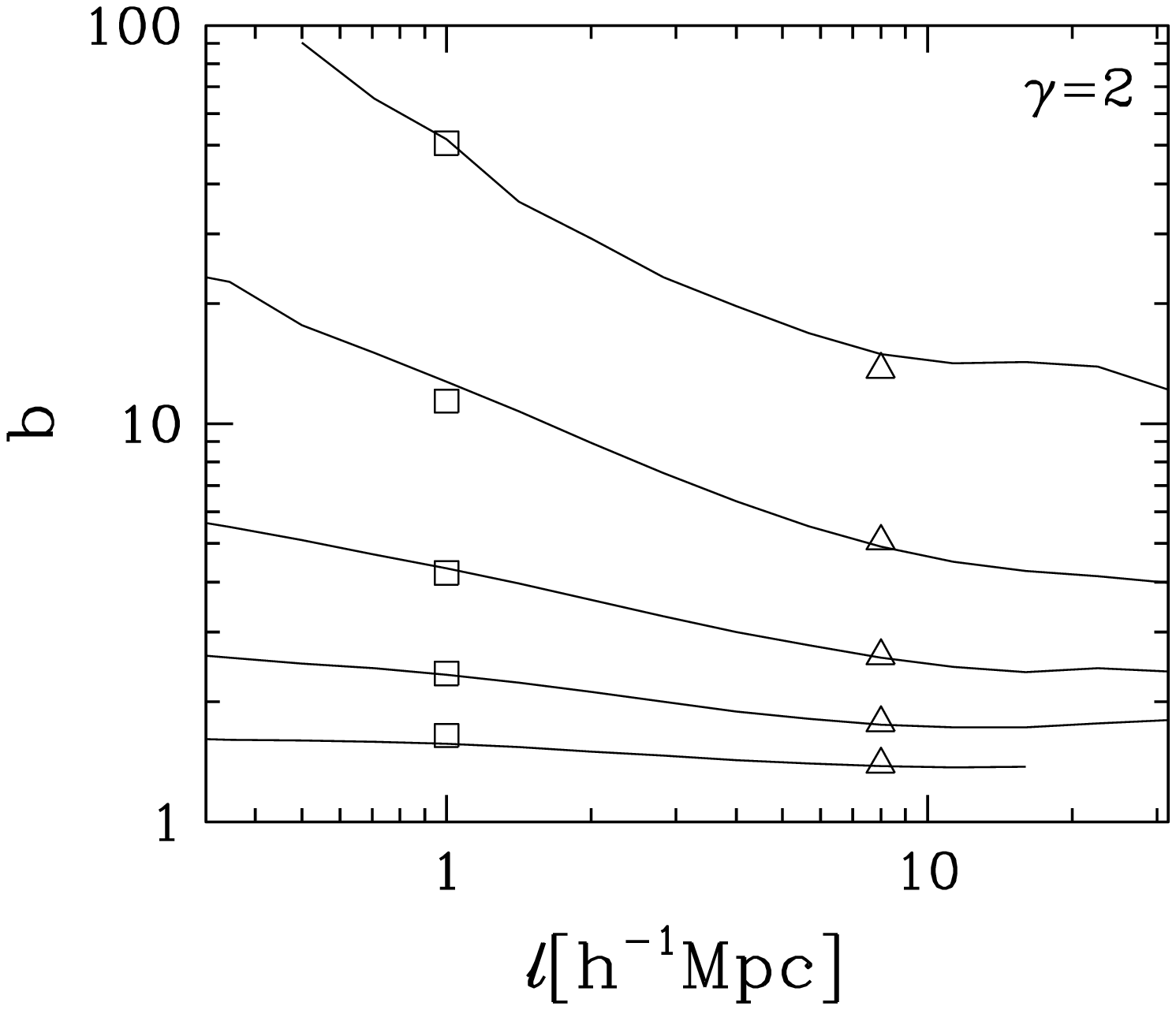}
\end{center}
\end{minipage}
\caption{The biasing function ${\tilde b}$ [equation
(\ref{eq:btilde})] as a function of spatial scale $\ell$, measured in
the mock galaxy catalogues.  The top and bottom panels correspond
respectively to $\gamma=1$ and $\gamma=2$ (by construction, we exactly
have ${\tilde b}=1$ for $\gamma=0$). Each curve is for a fixed value
of $z$, namely $z\simeq 0.18$, $0.33$, $0.62$, $1.25$ and $2.70$ from
bottom to top of each panel.  On the top panel, the squares and the
triangle give the values  expected from $b(z)=1+z$. On the bottom
panel, the square and the triangle correspond respectively to
$b(z)=(1+z)^3$ and $b(z)=(1+z)^2$.}
\label{fig:bias}
\end{center}
\end{figure}

Figure~\ref{fig:bias} shows the   scaling behavior of the bias factor
defined by
\begin{equation}
{\tilde b}(z,\ell)\equiv \frac{\sigma_{\rm gal.}(\ell,z)}{\sigma(\ell,z)},
\label{eq:btilde}
\end{equation}
as measured in the mock catalogues with $\gamma=1$ and $\gamma=2$.  In
this equation, $\sigma_{\rm gal.}^2(\ell)$ and $\sigma^2(\ell)$ are
respectively the variances in a sphere of radius $\ell$ of the galaxy
and the matter density distribution.  In fact, we take for $\sigma^2$
the variance measured in the mock catalogue with $\gamma=0$ which is
unbiased by definition.  To correct for variations of the selection
function we use the method proposed by Colombi, Szapudi \& Szalay
(1998). The curves on each panel correspond to redshift slices of
$[0.16,0.20]$, $[0.3,0.36]$, $[0.56,0.68]$, $[1.09,1.41]$ and
$[2.40,3.00]$.

By construction, the value of ${\tilde b}$ measured  at
$\ell=8^{-1}$Mpc (triangles) matches very well relation
(\ref{biasdef}). However there is no guarantee for this result to hold
at all scales. In other words, at fixed $z$, function ${\tilde
b}(\ell,z)$ is not necessarily a constant of scale [and equal to
$b(z)=(1+z)^\gamma$], although this is pretty much the case  for the
$\gamma=1$ mock catalogue.

For the $\gamma=2$ mock catalogue, function ${\tilde b}(\ell,z)$
presents large variations with scale, increasing with redshift. This
can be modeled as a varying effective ${\tilde \gamma}(\ell)$, for
example ${\tilde \gamma}\simeq 3$ for $\ell=1h^{-1}$ Mpc (squares on
bottom panel of Fig.~\ref{fig:bias}). While converting scales to
angles, more relevant to our analysis,
the modeling in terms of a function ${\tilde \gamma}(\theta)$ is not
very convincing. Still, we find that in the range of interest, $4 \la
\theta \la 20-30$ arcmin,  we should compare measured SLC effects to
semi-analytic predictions corresponding to $2 \la \gamma \la 3$.

\bsp 
\label{lastpage}

\end{document}